\documentclass[12pt]{article}

\usepackage{amsmath,amssymb,mathtools,mathrsfs} %数学記号
\usepackage[italicdiff]{physics}                %物理数学便利コマンド
\usepackage{type1cm}
\usepackage[T1]{fontenc}
\usepackage{fancyhdr}
\usepackage{appendix}
\usepackage{titlesec}
\usepackage{cite}
\usepackage{color}
\usepackage[top=25truemm,bottom=25truemm,left=20truemm,right=20truemm]{geometry}
\usepackage[dvipdfmx]{graphicx}
\usepackage{color}
\allowdisplaybreaks[4]

\baselineskip=\normalbaselineskip

\newcommand\als[1]{\begin{align}\begin{split}#1\end{split}\end{align}}

\makeatletter
 
 \@addtoreset{equation}{section}
\makeatother

\begin{document}
\setcounter{footnote}{0}
\setcounter{tocdepth}{3}
\bigskip
\def\thefootnote{\arabic{footnote}}

%%%%%%%%%%%%%%%%%%%%%%%%%%%%%%%%%%%%%%%%%%%%%%%%%%%
\begin{titlepage}
\renewcommand{\thefootnote}{\fnsymbol{footnote}}
\begin{normalsize}
\begin{flushright}
\begin{tabular}{l}
UTHEP-784
%DIAS-STP-21-22
\end{tabular}
\end{flushright}
  \end{normalsize}

~~\\

\vspace*{0cm}
    \begin{Large}
%    \begin{bf}
       \begin{center}
         {Matrix regularization for gauge theories}
       \end{center}
%    \end{bf}   
    \end{Large}

\vspace{0.7cm}

\begin{center}
Hiroyuki A\textsc{dachi}\footnote[1]
            {
e-mail address : 
s2030033@u.tsukuba.ac.jp},
Goro I\textsc{shiki}$^{1),2)}$\footnote[2]
            {
e-mail address : 
ishiki@het.ph.tsukuba.ac.jp},
Satoshi K\textsc{anno}$^{1)}$\footnote[3]
            {
e-mail address : 
kanno@het.ph.tsukuba.ac.jp}

\vspace{0.7cm}

     $^{ 1)}$ {\it Graduate School of Science and Technology, University of Tsukuba, }\\
               {\it Tsukuba, Ibaraki 305-8571, Japan}\\

     $^{ 2)}$ {\it Tomonaga Center for the History of the Universe, University of Tsukuba, }\\
               {\it Tsukuba, Ibaraki 305-8571, Japan}\\
               \end{center}

\vspace{0.5cm}

\begin{abstract}
\noindent
We consider how gauge theories can be described by matrix models. 
Conventional matrix regularization is defined for scalar functions and is not applicable to gauge fields, which are connections of fiber bundles. We clarify how the degrees of freedom of gauge fields are related to the matrix degrees of freedom, by formulating the Seiberg-Witten map between them.
\end{abstract}

\end{titlepage}

\tableofcontents

%%%%%%%%%%%%
%%%%%%%%%%%%
\section{Introduction}
%\hspace{0.51cm}
%%%%%%%%%%%%
%%%%%%%%%%%%
The matrix models are expected to describe superstring theory and M-theory\cite{Banks:1996vh,Ishibashi:1996xs,Dijkgraaf:1997vv}, and are thought to be able to describe various degrees of freedom of
extended objects in those theories such as strings, D-branes and M2/M5-branes.  
In particular, for coincident D-branes, non-Abelian gauge fields arise as massless modes of open strings, and the matrix models should also be able to 
describe gauge theories. 
In fact, many concrete examples 
\cite{Carow-Watamura:1998zks,Aoki:1999vr,Iso:2001mg,Kimura:2001uk,Steinacker:2003sd,Ishii:2007sy,Ishii:2008ib,Ishiki:2008vf,Ishiki:2009vr,Ishiki:2010pe,Fatollahi:2000wg,Fatollahi:2015xzf}
have shown that this is indeed possible.
Although matrices and gauge fields are mathematically quite different (The former is just finite sets of numbers with a specific algebraic 
structure, while the latter is connection 1-forms of fiber bundles), 
there are certainly several known ways to link them together. 

When the space-time is flat, there is a known map called the Seiberg-Witten map
\cite{Seiberg:1999vs}, 
which relates the ordinary gauge field to a noncommutative gauge field which has a deformed gauge transformation law 
with the star product. 
Since the star product is induced from the operator product, 
the noncommutative gauge field can be regarded as a field that takes the value in operators, or in other words, matrices of infinite dimension. Thus, the Seiberg-Witten map provides an example of relating a gauge field to a matrix. 
%Some extensions 
%of this map for non-flat spaces are also discussed in [**].

Such correspondences between gauge theories and matrix models are also 
known when the space is compact and the matrix size is finite. 
Theories on sphere and torus
\cite{Carow-Watamura:1998zks,Iso:2001mg,Kimura:2001uk}
are the best understood concrete examples.
The gauge fields arise as tangential fluctuations of  
the corresponding noncommutative background of the matrix models.
In these cases, by using the Fourier mode expansion
(as found in \cite{Ishii:2008ib} for example),  one can directly relates the degrees of 
freedom in gauge theories and those in matrix models in the large-$N$ limit.

Yet another method of formulating gauge fields from matrices is known
in the context of tachyon condensation
\cite{Asakawa:2001vm,Asakawa:2002ui,Terashima:2005ic}.
This method is closely related to a quantization scheme called the Berezin-Toeplitz quantization \cite{Bordemann:1993zv}, which is also discussed in this paper.
A gauge field appears as the connection 1-form used in this quantization and 
this field is identified with the dynamical gauge field on stable D-branes arising 
from the tachyon condensation.

There is also another formulation called Eguchi-Kawai model \cite{Eguchi:1982nm}, 
which was discovered earlier than the above examples.
The model is defined in terms of unitary matrices and is shown to 
reproduce the planar sector of a corresponding gauge theory 
provided that some modifications called quanching or twisting 
\cite{Bhanot:1982sh,Parisi:1982gp,Gross:1982at,Gonzalez-Arroyo:1982hyq}
is implemented, which protects the model from the notorious problem of the spontenious $U(1)^D$ symmetry breaking.

Thus, there are indeed several different methods of relating gauge fields and matrices.
However, these methods look quite different form each other and there seems 
to be a potential viewpoint that gives a unified understanding of them.
In this paper, we reconsider the problem of relating gauge fields and matrices
and find a new method  based on the Berezin-Toeplitz quantizaiton.
Our method also provides a relation between the above known methods
and gives a unifying understanding for this problem. 
By using this method, we can formulate the Seiberg-Witten map for finite size matrices.

The Berezin-Toeplitz quantization was also used
in the context of regularizing the worldvolume theory of a membrane
\cite{thesis-Hoppe} (The quantization is sometimes called matrix regularization 
in this context.).
The quantization is a map from continuous functions on the spatial world volume $M$
to a finite size matrices, such that the Poisson algebra of functions is 
semi-classically realized in the commutator algebra of matrices.
More specifically, for any function $f$ on $M$, its image $T(f)$ of 
the quantization map $T$ is a finite size (say $N\times N$) matrix and 
one of the main properties of the map is that 
$[T(f), T(g)]=-\frac{i}{N}T(\{f,g\}) +O(1/N^2)$ for any functions $f$ and $g$, 
where $\{\; , \; \}$ is a Poisson bracket on $M$.
For an embedding function $X^A: M \rightarrow R^D (A=1,2, \cdots, D)$ 
of the membrane living in 
the flat space, the quantization maps $X^A$ to a matrix $\hat{X}^A=T(X^A)$.
The infinitesimal area-preserving diffeomorphism of the form 
$\delta X^A=\{\alpha, X^A \}$, where $\alpha$ is a local parameter, 
is mapped in the above sense to the infinitesimal unitary similarity transformation
of the matrix, $\delta \hat{X}^A=iN [ \hat{\alpha}, \hat{X}^A] $. 
Thus, the regularized theory keeps the matrix version of the 
area-preserving diffeomorphism \cite{thesis-Hoppe}.

In this paper, we mainly focus on coincident D-branes in string theory 
and consider how they are described in terms of matrices. 
The data needed to describe the D-branes consist of the embedding function
$X$, which defines the position of the D-branes in the target space, and 
the excitations of open strings. In the low energy limit, 
the bosonic excitatons consists of a non-Abelian gauge field 
$B$ and adjoint scalar fields $\Phi$\footnote{In this paper, we use the letter 
$B$ for the non-Abelian gauge field. This should not be confused with the $B$-field 
in string theory which we do not deal with here.}. 
We construct a map from the data $(X, B, \Phi)$ to matrices 
$\hat{X}(X, B, \Phi)$, such that the gauge symmetry for $B$ and $\Phi$
is also realized in the commutator algebra of matrices.
Schematically, the map we will construct has the form 
\begin{align}
\hat{X}(X, B, \Phi)=T(X+\delta X(X,B,\Phi)),
\label{our result}
\end{align}
where $\delta X$ denote fluctuations of the embedding function $X$.
In the large-$N$ limit, the fields $B$ and $\Phi$ correspond to 
the tangential and vertical fluctuations of $X$, respectively. 
%The kind of this idea is found in \cite{Fatollahi:2000wg,Fatollahi:2015xzf} for example.
If we include the $1/N$ corrections, this relation 
becomes more complicated and we will argue that the relation is 
given by a generalization of the Seiberg-Witten map.
This relation works for arbitrary symplectic manifold $M$ and thus, gives 
a general way of relating gauge fields and matrices.
For $S^2$, we explicitly derive the Seiberg-Witten map up to the next leading 
order in $1/N$.

We then apply this method to a specific matrix model with a cubic interaction 
and relate the model with a gauge theory on $S^2$. 
The same relation was studied in the large-$N$ limit in 
\cite{Ishii:2007sy,Ishiki:2008vf} (See also \cite{Ishiki:2009vr,Ishiki:2010pe,Asano:2012gt}) in terms of the momentum expansion. 
Our method can also reveal the effect of $1/N$ corrections in 
this relation. We evaluate the $1/N$ correction of the classical action 
in this relation.

We also clarify the relation between our method and the other methods 
mentioned above.
We first show that the gauge field in \cite{Terashima:2005ic} can be identified
 in the large-$N$ limit with the gauge field $B$ in our setup. 
If we include $1/N$ corrections, they are related by a nonlinear 
redefinition of the fields. 
We then discuss a connection to Eguchi-Kawai model by
 applying the Berezin-Toeplitz quantization to Wilson line operators, not to the 
embedding function. The quantization maps Wilson lines to finite size matrices, and
in the case of torus, we demonstrate that the plaquete action written in terms 
of the Wilson lines is related to Eguchi-Kawai model through the quantization. 

This paper is organized as follows.
In section 2, we review the Berezin-Toeplitz quantization for vector bundles, 
which is needed in the subsequent section.
In section 3, we first review the Seiberg-Witten map and then generalize 
it for compact spaces based on the Berezin-Toeplitz quantization. Based on
this generalization, we find the map (\ref{our result}).
In section 4, we apply the Seiberg-Witten map to a cubic matrix model 
and see its relation to a massive BF theory on $S^2$.
In section 5, we discuss the relation between our method and the other 
formulations.
Section 6 is devoted to summary and discussion.

%%%%%%%%%%%%%%%%%%%%%%%%%%%%%%%%%%%%%%%%%%%%%%%%%%%%%%%%%%%%%%%%%%%%%
%%%%%%%%%%%%%%%%%%%%%%%%%%%%%%%%%%%%%%%%%%%%%%%%%%%%%%%%%%%%%%%%%%%%%
\section{Berezin-Toeplitz quantization for vector bundles}
\label{Berezin-Toeplitz quantization for vector bundles}
%%%%%%%%%%%%%%%%%%%%%%%%%%%%%%%%%%%%%%%%%%%%%%%%%%%%%%%%%%%%%%%%%%%%%
%%%%%%%%%%%%%%%%%%%%%%%%%%%%%%%%%%%%%%%%%%%%%%%%%%%%%%%%%%%%%%%%%%%%%
In this section, we review the Berezin-Toeplitz quantization for vector bundles
developed in 
\cite{Hawkins:1997gj,Hawkins:1998nj,Adachi:2020asg,Adachi:2021ljw,Adachi:2021aux}. For simplicity,
we consider 2-dimensional case, but higher dimensional case can also be treated in 
a similar manner \cite{Adachi:2022mln}.
See also \cite{Hasebe:2010vp,Nair:2020xzn,Hasebe:2023eiz}, where the quantization 
is discussed in terms of the Landau problem.

%%%%%%%%%%%%%%%%%%%%%%%%%%%%%%%%%%%%%%%%%%%%%%%%%%%%%%%%%%%%%%%%%%%%%%
%\subsection{Quantization of homomorphism bundles}
%%%%%%%%%%%%%%%%%%%%%%%%%%%%%%%%%%%%%%%%%%%%%%%%%%%%%%%%%%%%%%%%%%%%%

First, we describe the basic setup.
Let $M$ be a closed Riemann surface.
We denote by $g$ a Riemannian metric on $M$ and by $\omega$ 
the volume form of $g$. Since $\omega$ is a closed nondegenerate 2-form,
it gives a symplectic structure on $M$.
We also introduce the so-called prequantum line bundle $L$ such that it is a complex line bundle and its curvature is proportional to the symplectic form as
\begin{align}
    F^{(L)}=dA^{(L)}=\frac{\omega}{V},
\end{align}
where $A^{(L)}$ is the connection 1-form and
$V=\int_M \frac{\omega}{2\pi}$ is the volume of $M$.
Note that by definition the bundle $L$ has the unit monopole charge (the first Chern number) as
\begin{align}
    \frac{1}{2\pi}\int_M F^{(L)}=1.
\end{align}
We also introduce spinor fields on $M$ to construct the quantization. 
Let $S$ be the standard (2-component) spinor bundle on $M$. We then consider 
twisted spinor fields which are smooth sections of $S\otimes L^{p} \otimes E$, where
$p$ is a positive integer and $E$ is an arbitrary finite-rank vector bundle with hermitian metric and hermitian connection.
We define an inner product on $\Gamma(S\otimes L^{p} \otimes E)$ by
\footnote{In this paper, we denote by $\Gamma(E)$ a set of all smooth sections of a vector bundle $E$.}
\begin{align}
    (\psi',\psi):=\int_M \omega \psi'^{\dagger}\cdot \psi.
\end{align}
Here, $\cdot$ is a hermitian inner product on the fiber 
$(S\otimes L^{p} \otimes E)_x$ at each point $x\in M$.
The Dirac operator on $\Gamma(S\otimes L^{p} \otimes E)$ is defined in terms 
of local coordinates as
\begin{align}
    D^{(E)}\psi=i\gamma^{\alpha}\nabla^{(E)}_{\alpha}\psi.
\end{align}
Here, $\gamma^{\alpha}$ are the gamma matrices satisfying $\{\gamma^{\alpha},\gamma^{\beta}\}=g^{\alpha\beta}$.
% can rewrite the $\gamma^{\alpha}=\gamma^{a}e^{\alpha}_a$ by using the vierbein $e^{\alpha}_a$.
The covariant derivative $\nabla_\alpha^{(E)}$ acting on 
$\Gamma(S\otimes L^{p} \otimes E)$ is written as
\begin{align}
    \nabla_\alpha=\partial_\alpha + \Omega_\alpha -ipA^{(L)}_\alpha -iA_\alpha^{(E)},
\end{align}
where $\Omega_\alpha$ is the spin connection and $A^{(E)}_\alpha$ is the 
connection of $E$.
For sufficiently large $p$, the dimension of normalizable zero modes of the Dirac operator is given as 
${\rm dimKer}D^{(E)}=d^{(E)}p+c^{(E)}$,
where $d^{(E)}$ and $c^{(E)}$ are the rank and the first Chern number of $E$.
This follows from the Atiyah-Singer index theorem and the vanishing theorem \cite{Hawkins:1997gj,Hawkins:1998nj,Adachi:2021ljw}. 
We denote the projection operator onto the Dirac zero modes by 
$\Pi^{(E)} : \Gamma(S\otimes L^p \otimes E)\to {\rm Ker}D^{(E)}$.

Let $E$ and $E'$ be arbitrary finite-rank complex vector bundles over $M$ with 
hermitian inner products and hermitian connections.
Below, we focus on the homomorphism bundle ${\rm Hom}(E,E')$, which 
is a vector bundle whose fiber at a point $x\in M$ is a vector space of linear maps from the fiber $E_x$ to $E'_x$. 
The quantization we will construct below is a map from smooth 
sections of ${\rm Hom}(E,E')$ to finite-size matrices, that approximately 
preserves the algebraic structure of homomorphisms in the large matrix size limit.

Let us consider a field $\varphi \in \Gamma({\rm Hom}(E,E'))$. 
We can regard $\varphi$ as a linear map from
$\Gamma(S\otimes L^{p} \otimes E)$ to $\Gamma(S\otimes L^{p} \otimes E')$.
We define the quantization map for $\varphi \in \Gamma({\rm Hom}(E,E'))$ by
\footnote{Note that we make implicit the $E$- and $E'$- dependences on 
the left-hand side of  (\ref{quantization map}) to simplify the notation.}
\begin{align}
    T_p(\varphi):=\Pi^{(E')}\varphi\Pi^{(E)}.
    \label{quantization map}
\end{align}
The operator $T_p(\varphi)$ is called the Toeplitz operator of $\varphi$.
From the above estimation of ${\rm dimKer}D^{(E)}$, 
one finds that $T_p(\varphi)$ can be represented as a 
$(d^{(E')}p+c^{(E')})\times(d^{(E)}p+c^{(E)})$
 rectangular matrix. From the definition, it satisfies
\begin{align}
    T_p(\varphi^{\dagger})=T_p(\varphi)^{\dagger},
\end{align}
where $\dagger$ on the left-hand side is the hermitian conjugate of the homomorphism, while that on the right-hand side is the hermitian conjugate
of the rectangular matrix with respect to the Frobenius inner product.

The quantization map (\ref{quantization map}) has some desired properties,
which follow from an asymptotic expansion of the Toeplitz operator
in the large-$p$ limit. Let us consider two fields, $\varphi \in \Gamma({\rm Hom}(E,E'))$ and $\varphi' \in \Gamma({\rm Hom}(E',E''))$.
For $T_p(\varphi)=\Pi^{(E')}\varphi\Pi^{(E)}$ and 
$T_p(\varphi')=\Pi^{(E'')}\varphi'\Pi^{(E')}$,
their product $T_p(\varphi')T_p(\varphi)$ satisfies
the following asymptotic expansion in $\hbar_p = V/p$ \cite{Adachi:2021ljw}:
\begin{align}
    T_p (\varphi')T_p (\varphi) = \sum_{s=0}^{\infty} \hbar_{p}^s T_p (C_s(\varphi',\varphi)).
    \label{asym exp}
\end{align}
Here, the symbols $C_s$ on the right hand side are the bilinear differential operators $C_s: \Gamma({\rm Hom}(E',E'')) \otimes \Gamma({\rm Hom}(E,E'))
\rightarrow \Gamma({\rm Hom}(E,E''))$. The first four symbols are explicitly given by
\begin{align}
C_0(\varphi',\varphi) &= \varphi'\varphi,\nonumber\\
C_1(\varphi',\varphi) &= -\frac{1}{2} (g^{\alpha\beta} +i W^{\alpha\beta})(\nabla_{\alpha}\varphi')(\nabla_{\beta}\varphi),\nonumber\\
C_{2}(\varphi',\varphi)&=\frac{1}{8}(g^{\alpha\beta} +i W^{\alpha\beta})(\nabla_{\alpha}\varphi')
(R+4F^{(E')}) (\nabla_{\beta}\varphi) \nonumber\\
& \quad +\frac{1}{8}(g^{\alpha\beta} +i W^{\alpha\beta})(g^{\gamma\delta} +i W^{\gamma\delta})
(\nabla_{\alpha}\nabla_{\gamma}\varphi')(\nabla_{\beta}\nabla_{\delta}\varphi), \nonumber\\
C_3(\varphi',\varphi)&=-\frac{1}{32}\left(g^{\alpha\beta}+iW^{\alpha\beta}\right)(\nabla_\alpha\varphi') (4F^{(E^{'})}+R)^2(\nabla_\beta\varphi)\nonumber\\
    &-\frac{1}{32}\left(g^{\alpha\beta}+iW^{\alpha\beta}\right)\left(g^{\gamma\delta}+iW^{\gamma\delta}\right)\left[(\nabla_\alpha\nabla_\gamma\varphi')(4F^{(E^{'})}+R)(\nabla_\beta\nabla_\delta\varphi)\right.\nonumber\\
    &\left.\quad\quad\quad\quad\quad\quad\quad\quad\quad\quad\quad\quad\quad\quad\quad\quad+(\nabla_\gamma\varphi')(\nabla_\alpha(4F^{(E^{'})}+R))(\nabla_\beta\nabla_\delta\varphi)\right]\nonumber\\
    &-\frac{1}{32}\left(g^{\alpha\beta}+iW^{\alpha\beta}\right)\left(g^{\gamma\delta}+iW^{\gamma\delta}\right)(\nabla_\alpha\nabla_\gamma\varphi')(\nabla_\beta(4F^{(E^{'})}+R))(\nabla_\delta\varphi)\nonumber\\
    &-\frac{1}{16}\left(g^{\alpha\beta}+iW^{\alpha\beta}\right)\left(g^{\gamma\delta}+iW^{\gamma\delta}\right)(\nabla_\alpha\nabla_\gamma\varphi')(2F^{(E^{'})}+R)(\nabla_\beta\nabla_\delta\varphi)\nonumber\\
    &-\frac{1}{48}\left(g^{\alpha\beta}+iW^{\alpha\beta}\right)\left(g^{\gamma\delta}+iW^{\gamma\delta}\right)\left(g^{\epsilon f}+iW^{\epsilon f}\right)(\nabla_\alpha\nabla_\gamma\nabla_\epsilon\varphi')(\nabla_\beta\nabla_\delta\nabla_f\varphi).
    \label{c0toc3}
\end{align}
Here, $W^{\alpha\beta}$ is the Poisson tensor induced from the symplectic form $\omega$, $F^{(E')}$ is defined by 
$F^{(E')} = \frac{1}{2} W^{\alpha \beta} F^{(E')}_{\alpha \beta}$ and
$R$ is the Ricci scalar for the metric $g$.
In terms of the inverse of the vierbein, $e_a^\alpha (a=1,2)$ satisfying 
$e_a^\alpha e_a^\beta = g^{\alpha \beta}$, the Poisson tensor is 
explicitly given as $W^{\alpha\beta}=e_1^\alpha e_2^\beta -e_2^\alpha e_1^\beta$.
Similarly, we have $F^{(E')} = e_1^\alpha e_2^\beta F_{\alpha \beta}^{(E')} =
e_1^\alpha e_2^\beta (\partial_\alpha A^{(E')}_\beta -\partial_\beta A^{(E')}_\alpha
-i[A^{(E')}_\alpha, A^{(E')}_\beta])$. For $\varphi \in \Gamma({\rm Hom}(E,E'))$ and 
$\varphi' \in \Gamma({\rm Hom}(E',E''))$, the covariant derivatives in 
(\ref{c0toc3}) are defined as
\begin{align}
    \nabla_\alpha \varphi = \partial_\alpha \varphi -iA_\alpha^{(E')} \varphi +i \varphi A_\alpha^{(E)},
\;\;\;
\nabla_\alpha \varphi' = \partial_\alpha \varphi' -iA_\alpha^{(E'')} \varphi' +i \varphi' A_\alpha^{(E')}.
\end{align}

If we use the local orthnormal basis defining 
$\nabla_{\pm}:=\frac{1}{\sqrt{2}}(\nabla_1\pm i\nabla_2)$, where 
$\nabla_a := e_a^\alpha \nabla_\alpha (a=1,2)$, the symbols 
$C_s$ can be written as
\begin{align}
C_0(\varphi',\varphi) &= \varphi'\varphi,\nonumber\\
C_1(\varphi',\varphi) &= -(\nabla_{-}\varphi')(\nabla_{+}\varphi ),\nonumber\\
C_{2}(\varphi',\varphi)&=\frac{1}{4}(\nabla_{-}\varphi' )
(R+4F^{(E')})( \nabla_{+}\varphi )+\frac{1}{2}(\nabla^2_{-}\varphi')
(\nabla^2_{+}\varphi), \nonumber\\
C_3(\varphi',\varphi)&=-\frac{1}{6}(\nabla^3_-\varphi') (\nabla^3_+\varphi)-\frac{1}{4}(\nabla^2_-\varphi') (4F^{(E^{'})}+\frac{3}{2}R)(\nabla^2_+\varphi)\nonumber\\
&\quad -\frac{1}{4}(\nabla_-\varphi') (\nabla_-(2F^{(E^{'})}+\frac{1}{2}R))(\nabla^2_+\varphi)
-\frac{1}{4}(\nabla^2_-\varphi') (\nabla_+(2F^{(E^{'})}+\frac{1}{2}R))(\nabla_+\varphi)\nonumber\\
&\quad -\frac{1}{16}(\nabla_-\varphi') (4F^{(E^{'})}+R)^2(\nabla_+\varphi).
\label{plusminus notation}
\end{align}

Note that when all of the vector bundles $E, E', E''$ and so on are the trivial line bundles, 
the above quantization reduces to the conventional matrix regularization for 
functions on $M$. 

For higher dimensional symplectic spaces, we can also define the quantization map 
satisfying an asymptotic relation, which takes the same form as 
(\ref{asym exp}) \cite{Adachi:2022mln}. 
In this case, the form of the bilinear differential operators $C_s$ 
are slightly modified from the 2-dimensional case
except $C_0$ and $C_1$, which remain to have the same form as above.

%%%%%%%%%%%%%%%%%%%%%%%%%%%%%%%%%%%%%%%%%%%%%%%%%%%%%%%%%%%%%%%%%%%%%
%%%%%%%%%%%%%%%%%%%%%%%%%%%%%%%%%%%%%%%%%%%%%%%%%%%%%%%%%%%%%%%%%%%%%
\section{Seibeg-Witten map for finite size matrices}
%%%%%%%%%%%%%%%%%%%%%%%%%%%%%%%%%%%%%%%%%%%%%%%%%%%%%%%%%%%%%%%%%%%%%
%%%%%%%%%%%%%%%%%%%%%%%%%%%%%%%%%%%%%%%%%%%%%%%%%%%%%%%%%%%%%%%%%%%%%
In this section, we first review the Seiberg-Witten map (SW map) for 
noncommutative plane \cite{Seiberg:1999vs} and then construct
the map (\ref{our result})
by generalizing the SW map for 
closed noncommutative manifolds described by finite-size matrices.
We also explicitly evaluate the SW map for $S^2$ up to the next leading order 
in $1/N$.

\subsection{Review of the Seiberg-Witten map on flat space}
Let us consider the gauge group $U(k)$ on a flat space.
We can consider two kinds of $U(k)$ gauge fields: one is the ordinary non-abelian 
gauge field $B$ which transforms as 
\begin{align}
    \delta_\beta B_a=\partial_a \beta -i[\beta, B_a],
    \label{delA}
\end{align}
and the other is the non-commutative gauge field $\hat{B}$ which has the deformed transformation law,
\begin{align}
    \delta_{\hat{\beta}} \hat{B}_a=\partial_a \hat{\beta}-i\hat{\beta}*\hat{B}_a
    +i\hat{B}_a*\hat{\beta}.
    \label{delAA}
\end{align}
Here, $*$ is the non-commutative star product (Moyal product) satisfying
\begin{align}
\hat{f} * \hat{g} = \hat{f}\hat{g} + O(\hbar), \;\;\;\;
\hat{f} * \hat{g} -\hat{g} * \hat{f} = -i \hbar \{\hat{f}, \hat{g} \} +O(\hbar^2)
\end{align}
for any functions $\hat{f}$ and $\hat{g}$,
where $\hbar$ corresponds to the non-commutative parameter and 
$\{ \; ,\; \}$ is the standard Poisson bracket on plane.

The SW map is a map from the ordinary gauge field to the non-commutative 
gauge field, 
\begin{align}
    \hat{B}=\hat{B}(B),\quad\quad \hat{\beta}=\hat{\beta}(\beta, B),
\end{align}
which is compatible with the above transformations \cite{Seiberg:1999vs}.
The compatibility of the transformations is expressed as 
\begin{align}
    \hat{B}(B)+ \delta_{\hat{\beta}} \hat{B}(B)=\hat{B}(B+\delta_\beta B).
    \label{compatiblity of sw map}
\end{align}
Assuming that the map has a power series expansion with respect to $\hbar$ as 
\begin{align}
   \hat{B} =B + O(\hbar), \quad\quad \hat{\beta}= \beta + O(\hbar),
\end{align}
the condition (\ref{compatiblity of sw map}) can be pertrubatively solved. 
The solution at the first order in $\hbar$ is given as 
\begin{align}
   \hat{B}_a =B_a -\frac{\hbar}{4}\epsilon^{bc}
   \{B_b, \partial_c B_a +F_{ca}^0 \}
    + O(\hbar^2), \quad\quad 
    \hat{\beta}= \beta + \frac{\hbar}{4}\epsilon^{ab}\{\partial_a \beta, B_b \}  +O(\hbar^2),
    \label{original SW map solution}
\end{align}
where $F_{ab}^0 = \partial_a B_b -\partial_b B_a -i[B_a, B_b]$ is the 
field strength of $B$.

%%%%%%%%%%%%%%%%%%%%%%%%%%%%%%%%%%%%%%%%%%%%%%%%%%%%%%
\subsection{The Seiberg-Witten map for finite-size matrices}
\label{The Seiberg-Witten map for finite-size matrices}
%%%%%%%%%%%%%%%%%%%%%%%%%%%%%%%%%%%%%%%%%%%%%%%%%%%%%%
  Let us consider the $U(k)$ gauge group on a closed symplectic space $M$. 
 Let $B_a$ and $\chi_i$ be the gauge field and some adjoint scalar fields on $M$, 
which transform as
\begin{align}
    \delta_\beta B_a=\partial_a \beta+i[\beta,B_a],\quad \delta_\beta \chi_i=i[\beta,\chi_i].
\end{align}
Below, 
for a given isometric embedding $X^A (A=1,2, \cdots, D) : M \rightarrow R^D $,
we will define a map from the data $(X^A, B_a, \chi_i )$ to 
finite size matrices $\hat{X}^A$ such that the above gauge transformation 
is compatible with the unitary similarity transformation of $\hat{X}^A$ generated by
\begin{align}
    \delta_{\hat{\alpha}} \hat{X}^A=i[\hat{\alpha} ,\hat{X}^A].
    \label{D0 transformation}
\end{align}
The compatibility condition   
 is expressed as
\begin{align}
\hat{X}^A(X, B, \chi) + \delta_{\hat{\alpha}} \hat{X}^A(X, B, \chi)
=\hat{X}^A(X, B+\delta_{\beta}B, \chi + \delta_{\beta}\chi),
\label{compatibility of xhat}
\end{align}
which is very similar to (\ref{compatiblity of sw map}).
Thus, the map can be seen as a generalization of the SW map.

Our ansatz for the map is
\begin{align}
    \hat{X}^A&=T_p(X^A1_k+\hbar_p \Tilde{X}^A(X,B,\chi)), \nonumber\\
    \hat{\alpha} &=T_p (\alpha (X,B,\chi, \beta))
    \label{hatX}
\end{align}
where, on the right-hand side,
the Berezin-Toeplitz quantization is defined for an endmorphism bundle 
(namely, ${\rm Hom}(E,E')$ with $E=E'$) on 
the $k$-dimensional vector bundle $E$ with connection $A^{(E)}$ and 
$1_k$ is the identity matrix on the fiber of $E$.
$\Tilde{X}^A$ and $\alpha$ take values in $\Gamma({\rm Hom}(E,E))$ and 
$\Tilde{X}^A$ can be considered as fluctuations $X^A$.
The tortal matrix size of $\hat{X}^A$ is given by $N=kp+c^{(E)}$.
This ansatz is very natural from the viewpoint of D-branes, since 
the gauge field and scalar fields are known to arise as tangential and 
vertical fluctuations of D-branes, respectively.
Taking this analogy further, we decompose the fluctuation to 
the tangential and vertical directions as
\begin{align}
    \Tilde{X}^A(X,B,\chi)=W^{ab}A_a(X,B,\chi) \nabla_bX^A+\phi_i (X,B,\chi) Y_i^A.
    \label{expantion of fluctuations}
\end{align}
Here, $W$ is the Poisson tensor of $M$ and
 $a,b$ are indices of an orthonormal basis such that
$\nabla_a X^A \nabla_b X^A = \delta_{ab}$ and 
$Y_i^A$ are orthonormal vectors perpendicular to $\nabla_bX^A$ satisfying
$Y_j^AY_i^A =\delta_{ij}$ and $Y_i^A\nabla_bX^A =0$ for every point on $M$\footnote{Recall that we have assumed that $X$  is isometric embedding and hence 
satisfies $\partial_\alpha X^A \partial_\beta X^A = g_{ab}$, 
where $g$ is a metric 
on $M$. Multiplying the vielbeins, we obtain $\nabla_aX^A \nabla_bX^A = \delta_{ab}$
in the orthonormal basis.}.
Note that when $M$ is $2n$-dimensional space, 
$a, b$ run from $1$ to $2n$ while $i$ runs from $1$ to $D-2n$ 
and the vectors $\{\nabla_a \vec{X}, \vec{Y}_i\}$ 
form an orthonormal basis of the $D$-dimensional vector space.
The fields $A_a$ and $\phi_i$ take values in $\Gamma({\rm Hom}(E,E))$ and 
hence they can be represented as $k \times k$ matrices.
From the perspective of D-branes, $A_a$ and $\phi_i$ seem to directly 
correspond to $B_a$ and $\chi_i$, respectively.
As we will see below, this is indeed correct
in the classical limit $\hbar_{p}\rightarrow 0$.
However, when $\hbar_{p}$ corrections are taken into account, 
the compatibility condition (\ref{compatibility of xhat}) implies more complicated 
nonlinear relations between them, which are very similar to the SW map
(\ref{original SW map solution}).

By substituting our ansatz (\ref{hatX}) into the transformation law
(\ref{D0 transformation}), 
and using the asymptotic expansion (\ref{asym exp}),
we find that (\ref{D0 transformation}) implies the following transformation laws for $A_a$ and $\phi_i$\footnote{Here, since our ansatz (\ref{hatX}) 
relates $\hat{\alpha}$ with $\alpha$, we write $\delta_{\hat{\alpha}}$ simply 
as $\delta_\alpha$.},
\begin{align}
    \label{eq:trans_A}
    \delta_\alpha A_a&=i\sum_{s=1}^{\infty}\hbar_p^{s-1}W_{ab}g^{bc}\nabla_c X^A(D_s(\alpha,X^A1_k)+D_{s-1}(\alpha,\Tilde{X}^A))\nonumber\\
    \delta_\alpha \phi_i&=i\sum_{s=1}^{\infty}\hbar_p^{s-1}Y_i^A(D_s(\alpha,X^A1_k)+D_{s-1}(\alpha,\Tilde{X}^A)),
\end{align}
where, $D_s(\varphi,\phi)=C_s(\varphi,\phi)-C_s(\phi,\varphi)$.
In the classical limit $\hbar_{p}\rightarrow 0$, (\ref{eq:trans_A}) reduces to 
\begin{align}
    \delta_\alpha A_a  &=\partial_a \alpha+i[\alpha,A_a+A^{(E)}_a]+O(\hbar_p),\nonumber\\
    \delta_\alpha \phi_i&=i[\alpha,\phi_i]+O(\hbar_p).
\end{align}
These are the transformation laws of non-Abelian gauge field and 
adjoint scalar field, if we interpret $A^{(E)}$ as the background of $A$. 
Thus, we can assume the following form for the solution of the compatibility
(\ref{compatibility of xhat}):
\begin{align}
    A_a(B,\chi)&=B_a-A^{(E)}_a +\hbar_p B_{1,a}(B,\chi)+\hbar_p^2 B_{2,a}(B,\chi)+\cdots, \nonumber\\
    \phi_i(B,\chi)&=\chi_i+\hbar_p \chi_{1,i}(B,\chi)+\hbar_p^2 \chi_{2,i}(B,\chi)+\cdots, \nonumber\\
    \alpha(B,\chi,\beta)&=\beta+\hbar_p \beta_{1}(B,\chi,\beta)+\hbar_p^2 \beta_2(B,\chi,\beta)+\cdots.
    \label{SW expansion}
\end{align}
The compatibility (\ref{compatibility of xhat})  is written in terms of 
$A_a$ and $\phi_i$ as 
\begin{align}
    \label{SWeq}
    A_a(B,\chi)+\delta_\alpha A_a(B,\chi)=A_a(B+\delta_\beta B,\chi+\delta_\beta \chi),\nonumber \\
    \phi_i(B,\chi)+\delta_\alpha \phi_i(B,\chi)=\phi_i(B+\delta_\beta B,\chi+\delta_\beta \chi),
\end{align}
where we omitted the $X$ dependence.
We can determine the higher order terms in (\ref{SW expansion})
by solving these conditions.
We will demonstrate this for $S^2$ below.

Note that though we do not see any local gauge symmetry 
on the matrix side (\ref{D0 transformation}), the above map relates 
such objects with the gauge field with local gauge symmetry.
The emergence of the local gauge symmetry 
originates from a symmetry that keeps $\hat{X}^A$ invariant.
In fact, under the following transformation,
\begin{align}
A^{(E)} &\rightarrow A^{U(E)}:= UA^{(E)}U^{-1}+ iUdU^{-1}, \nonumber\\
B &\rightarrow B^U:=UBU^{-1}+ iUdU^{-1}, \nonumber\\
\chi &\rightarrow \chi^U := U\chi U^{-1},
\label{gauge symmetry of Teoplitz operator}
\end{align}
the Toeplitz operator (\ref{hatX}) is invariant.
Thus, the gauge symmetry can be understood as the gauge redundancy 
of $A^{(E)}$ that constitutes the quantization scheme.
The existence of the above symmetry is shown in 
Appendix~\ref{The gauge symmetry of the Toeplitz operator}.

%%%%%%%%%%%%%%%%%%%%%%%%%%%%%%%%%%%%%%%%%%%%%%%%%%%%%%%%%%%%%
\subsection{The Seiberg-Witten map for fuzzy $S^2$}
\label{SW map for fuzzy S^2}
%%%%%%%%%%%%%%%%%%%%%%%%%%%%%%%%%%%%%%%%%%%%%%%%%%%%%%%%%%%%%
Here, we consider fuzzy $S^2$ and we find a solution to (\ref{SWeq}) up to the next-leading order in $\hbar_p$. 

Let us consider $S^2$ with the standard metric and its 
isometric embedding into $R^3$, satisfying 
\begin{align}
X^A X^A =1.
\label{xx=1}
\end{align}
In this case, the scalar curvature $R$ is equal to $2$ and
the vector perpendicular to $S^2$ is given by $Y^A=X^A$.
By using the notation $A_{\pm}=\frac{1}{\sqrt{2}}(A_1\pm iA_2)$
(see Appendix~\ref{The plus-minus notation} for this notation).
 (\ref{expantion of fluctuations}) is 
written as
\begin{align}
    \Tilde{X}^A&=iA_- \nabla_+X^A-iA_+ \nabla_-X^A+\phi X^A.
\end{align}
For simplicity, we assume that $A^{(E)}=0$.
Then, the transformation laws (\ref{eq:trans_A}) for the fluctuations are explicitly 
given as
\begin{align}
    \label{varia_Aphi}
    \delta_\alpha\phi&=i[\alpha,\phi]+\hbar_p(-A_-\nabla_+\alpha - \nabla_-\alpha A_+ + i\nabla_-\phi\nabla_+\alpha - i\nabla_-\alpha\nabla_+\phi) \nonumber\\
    &\quad +\hbar_p^2\left(+\frac{1}{2}A_-\nabla_+\alpha + \nabla_-A_-\nabla^2_+\alpha + \frac{1}{2}\nabla_-\alpha A_+ + \nabla^2_-\alpha\nabla_+A_+ \right.\nonumber\\
    &\quad\quad\quad\quad\quad \left. +i\frac{1}{2}\nabla_-\phi\nabla_+\alpha- i\frac{1}{2}\nabla_-\alpha\nabla_+\phi+\frac{i}{2}\nabla^2_-\phi\nabla^2_+\alpha-\frac{i}{2}\nabla^2_-\alpha\nabla^2_+\phi \right)+\cdots, \nonumber\\
    \delta_\alpha A_-&=\nabla_-\alpha + i[\alpha, A_-] + \hbar_p\qty(-\frac{1}{2}\nabla_-\alpha + i\nabla_-A_-\nabla_+\alpha - i\nabla_-\alpha\nabla_+A_- + \nabla_-\alpha\phi)\nonumber\\
    &\quad +\hbar^2_p\left( \frac{R}{16}\nabla_-\alpha + \frac{iR}{4}\nabla_-\alpha\nabla_+A_- + \frac{i}{2}\nabla^2_-\alpha\nabla^2_+A_- - \frac{iR}{4}\nabla_-A_-\nabla_+\alpha - \frac{i}{2}\nabla^2_-A_-\nabla^2_+\alpha \right. \nonumber\\
    &\quad\quad\quad\quad\quad\quad\quad\quad\quad\quad\quad\quad\quad\quad\quad\quad\quad\quad\quad\quad \left. +\frac{i}{2}\nabla^2_-\alpha A_+ - \frac{1}{2}\nabla_-\alpha\phi - \nabla^2_-\alpha\nabla_+\phi \right)+\cdots, \nonumber\\
    \delta_\alpha A_+&=\nabla_+\alpha + i[\alpha, A_+] + \hbar_p\qty(-\frac{1}{2}\nabla_+\alpha + i\nabla_-A_+\nabla_+\alpha - i\nabla_-\alpha\nabla_+A_+ + \phi\nabla_+\alpha) \nonumber\\
    &\quad +\hbar^2_p\left( \frac{R}{16}\nabla_+\alpha + \frac{iR}{4}\nabla_-\alpha\nabla_+A_+ + \frac{i}{2}\nabla^2_-\alpha\nabla^2_+A_+ - \frac{iR}{4}\nabla_-A_+\nabla_+\alpha - \frac{i}{2}\nabla^2_-A_+\nabla^2_+\alpha \right.\nonumber\\
    &\quad\quad\quad\quad\quad\quad\quad\quad\quad\quad\quad\quad\quad\quad\quad\quad\quad\quad\quad\quad \left. +\frac{i}{2}A_-\nabla^2_+\alpha - \frac{1}{2}\phi\nabla_+\alpha - \nabla_-\phi\nabla^2_+\alpha \right)+\cdots.
\end{align}
In deriving above equations, we used an equation shown in 
Appendix~\ref{A formula for the isometric embedding of S2}.
By integrating these equations, 
we can find that $\chi_1,B_{1,-},B_{1,+}$ and $\beta_1$ in (\ref{SW expansion})
are given as \footnote{Here, $\chi_1$ represents $\chi_{1,i}$ with $i=1$ 
in (\ref{SW expansion}).}
\begin{align}
    \label{S2SWmap}
    \chi_1&=-B_-B_+ + i\nabla_-\chi B_+ - iB_-\nabla_+\chi - \frac{1}{2}B_-[B_+ , \chi] + \frac{1}{2}[ B_-, \chi]B_+, \nonumber\\
    B_{1,-}&=-\frac{1}{2}B_- + B_-\chi +\frac{i}{2}(\nabla_-B_-)B_+ - iB_-\nabla_+B_- + \frac{i}{2}B_-\nabla_-B_+ + \frac{1}{2}B_-[B_-,B_+], \nonumber\\
    B_{1,+}&=-\frac{1}{2}B_+ + \chi B_+ -\frac{i}{2}B_-\nabla_+B_+ + i(\nabla_-B_+)B_+ - \frac{i}{2}(\nabla_+B_-)B_+ + \frac{1}{2}[B_-,B_+]B_+, \nonumber\\
    \beta_1&=-\frac{i}{2}B_-\nabla_+\beta + \frac{i}{2}\nabla_-\beta B_+.
\end{align}
A derivation of the above solution is shown in 
Appendix~\ref{The derivation of the SW map of S2}.

%%%%%%%%%%%%%%%%%%%%%%%%%%%%%%%%%%%%%%%%%%%%%%%%%%%%%%%%%%%%%%%%%%%%%
\section{Application to a cubic matrix model}
%%%%%%%%%%%%%%%%%%%%%%%%%%%%%%%%%%%%%%%%%%%%%%%%%%%%%%%%%%%%%%%%%%%%%
%%%%%%%%%%%%%%%%%%%%%%%%%%%%%%%%%%%%%%%%%%%%%%%%%%%%%%%%%%%%%%%%%%%%%
In this section, we apply the SW map in section \ref{SW map for fuzzy S^2} to 
a cubic matrix model,
\begin{align}
    S={\rm Tr}\qty(\hat{X}^A\hat{X}^A+\frac{i\alpha}{\hbar_p}\epsilon^{ABC}\hat{X}^A\hat{X}^B\hat{X}^C),
    \label{cubic MM}
\end{align}
and see how this model can be related to a gauge theory on $S^2$.
 Such relation was also studied in earlier literatures \cite{Ishii:2007sy,Ishiki:2008vf}
  from different view points, and the above model is known to correspond to a 
  massive BF theory on $S^2$ in the large-$N$ limit. 
The SW map we constructed above enables us to see this relation including 
the $1/N$ corrections. 
Below, we assume that $A^{(E)}=0$ for simplicity, and the 
covariant derivatives shall act as
\begin{align}
    \nabla_\alpha \varphi
    = \partial_\alpha \varphi,\quad
    \nabla_\alpha \xi_\beta
    = \partial_\alpha \xi_\beta + \Gamma^\gamma_{\alpha\beta}\xi_\gamma
\end{align}
for $\varphi \in \Gamma( {\rm Hom}(E,E))$ and $\xi_\alpha \in 
\Gamma({\rm Hom}(E,E)\otimes T^*M)$.
We write detailed computation in Appendix~\ref{The calculation for the qubic matrix model} and only present  the result here.

By using \eqref{S2SWmap}, the action (\ref{cubic MM}) 
is rewritten in terms of the non-Abelian gauge field $B$ and the adjoint scalar 
field $\chi$ as
\begin{align}
    S&=\int_M \frac{\omega}{4} {\rm tr}\left[
    2\chi 
    + \hbar_p(2F_{12}^0\chi -2\chi + \chi^2) 
    + \hbar_p^2\qty(-\chi^2 - D_-\chi D_+\chi + \chi^2F^0_{12} + \frac{R}{2}\chi)
    \right] \nonumber\\
    &\quad
    +\alpha\int_M \frac{\omega}{4} {\rm tr}\left[
    -3\chi -3\hbar_p\qty(\chi^2-\frac{3}{2}\chi)\right.
    -3\hbar^2_p\left(-\frac{3}{2}F^0_{12}\chi - \frac{1}{3}\chi^3 - \frac{3}{2}\chi^2 + \frac{5}{4}\chi \right.
    \nonumber\\
    &\left.\quad\quad\quad \left.
     - \frac{1}{2}(F^0_{12})^2 - 2D_-\chi D_+\chi +  \frac{1}{2}F^0_{12}(D_-D_+ + D_+D_-)\chi \right)
     \right]+ O(\hbar_p^3),
     \label{BF plus quantum corrections}
\end{align}
where we have defined the field strength of $B$  and the gauge covariant derivative 
as
\begin{align}
    F^0_{12}&=-i(\nabla_-B_+-\nabla_+B_- -i[B_-,B_+]), \nonumber\\
    D_{\pm}\chi&=\nabla_\pm\chi - i [B_\pm,\chi].
\end{align}
By fine-tuning the parameter $\alpha$ as $\alpha=\frac{2}{3} -\frac{1}{3}\hbar_p + \cdots$, 
one can always eliminate the linear term of $\chi$. Then, for small $\hbar_p$, 
we have
\begin{align}
    S=\int_M \frac{\omega}{2\pi}\hbar_p {\rm tr}(2F^0_{12}\chi -\chi^2)+O(\hbar_p^2).
\end{align}
This is indeed the action of the massive BF theory.
If we integrating out $\chi$, the above action reduces to 
the 2-dimensional pure Yang-Mills theory, which is known to be a 
topological field theory. It would be interesting to study how the 
$\hbar_p$-corrections in (\ref{BF plus quantum corrections}) 
modify the topological nature of the Yang-Mills theory.

A similar calculation is also possible for more general matrix models.
For example, for a model with quartic and cubic interactions, 
which was also studied earlier in \cite{Azuma:2004zq,Azuma:2004ie},
we have
\begin{align}
    &{\rm Tr}\qty(-\frac{1}{4}[\hat{X}^A,\hat{X}^B][\hat{X}^A,\hat{X}^B] + \frac{2}{3}i\alpha\hbar_p\epsilon^{ABC}\hat{X}^A \hat{X}^B \hat{X}^C) \nonumber\\
    =& \int \frac{\omega}{4}{\rm tr} \hbar_p^2 \left[
    2(1-\alpha)\chi
    + \hbar_p\qty(
    \frac{1}{2}(F_{12}^0)^2
    + D_-\chi D_+\chi
    + (3-2\alpha)\chi^2
    - 2F_{12}^0\chi
    + (3\alpha-1)\chi
    ) + O(\hbar_p^4)
    \right].
\end{align}

%%%%%%%%%%%%%%%%%%%%%%%%%%%%%%%%%%%%%%%%%%%%%%%%%%%%%%%%%%%%%%%%%%%%%
\section{Relations to other formulations}
%%%%%%%%%%%%%%%%%%%%%%%%%%%%%%%%%%%%%%%%%%%%%%%%%%%%%%%%%%%%%%%%%%%%%
%%%%%%%%%%%%%%%%%%%%%%%%%%%%%%%%%%%%%%%%%%%%%%%%%%%%%%%%%%%%%%%%%%%%%
There are other formulations of gauge fields in terms of matrices.
In this section, we will consider relations between our description and the others.

%%%%%%%%%%%%%%%%%%%%%%%%%%%%%%%%%%%%%%%%%%%%%%%%%%%%%%%%%%%%%%%%%%%%%
\subsection{Gauge fields in tachyon condensation}
%%%%%%%%%%%%%%%%%%%%%%%%%%%%%%%%%%%%%%%%%%%%%%%%%%%%%%%%%%%%%%%%%%%%%
It was shown in earlier work that the Berezin-Toeplitz 
quantization also plays important roles in the context of tachyon condensation
\cite{Terashima:2005ic}.
In this context, the connection 1-form $A^{(E)}$ in the Berezin-Toeplitz quantization 
is identified with the gauge field on D-branes. 

In the classical limit $\hbar_p\rightarrow0$,
one can identify $A^{(E)}$ with $B$, since they have the common transformation 
laws (\ref{gauge symmetry of Teoplitz operator}) and $A^{(E)}$ in (\ref{SW expansion}) can be eliminated by shifting $B$ as $B\rightarrow B+A^{(E)}$.
For finite $\hbar_p$,  $A^{(E)}$ in the higher order terms in (\ref{SW expansion}) cannot be eliminated by such a simple shift. 
Instead, we can consider $B \rightarrow B +A^{(E)} +O(\hbar_p)$ with the 
last term of $O(\hbar_p)$ suitably fine-tuned to eliminate $A^{(E)}$'s in 
(\ref{SW expansion}).
There may be a subtlety here
when the bundle $E$ has non-vanishing Chern numbers.
In this case, since the matrix size of the quantization map depends on the Chern number of $E$, it would not be possible to eliminate $A^{(E)}$'s keeping the 
same size of the matrices. Thus, above discussion should be 
modified such that only the fluctuations of $A^{(E)}$ that do not change 
the Chern numbers can actually be absorbed by shifting $B_a$.

%%%%%%%%%%%%%%%%%%%%%%%%%%%%%%%%%%%%%%%%%%%%%%%%%%%%%%%%%%%%%
%%%%%%%%%%%%%%%%%%%%%%%%%%%%%%%%%%%%%%%%%%%%%%%%%%%%%%%%%%%%%
\subsection{Matrix regularization for Wilson line operators}
%%%%%%%%%%%%%%%%%%%%%%%%%%%%%%%%%%%%%%%%%%%%%%%%%%%%%%%%%%%%%
%%%%%%%%%%%%%%%%%%%%%%%%%%%%%%%%%%%%%%%%%%%%%%%%%%%%%%%%%%%%%
In this section, we describe the matrix regularization for Wilson line operators.
We consider the 2-dimensional torus $T^2$ with the volume $V=2\pi$ for simplicity but it is easy to generalize it into the higher dimensional torus $T^{2n}$.

We introduce a new gauge field $A$ for the bundle $E$ that is 
independent of $A^{(E)}$, 
and consider the problem of quantizing $A$.
Below, we consider the Berezin-Toeplitz quantization with $A^{(E)}=0$ for simplicity, 
but $A$ can still be nontrivial\footnote{
If we consider the field $A_\alpha=A_\alpha^{(E)} + \Phi_\alpha (\alpha=1,2) $,
where $\Phi_\alpha$ are adjoint scalars, $A$ has the same transformation law 
with $A^{(E)}$. Even when $A^{(E)}=0$, $A$ is nontrivial by the degrees of freedom 
of $\Phi_\alpha$.}.
Since $A$ is a connection and the gauge transformation law of $A$ is different from that of local sections, the quantization in section 
\ref{Berezin-Toeplitz quantization for vector bundles}
is not applicable to $A$. 
Instead, let us consider the following operators made of Wilson lines of $A$
and acting on $\psi \in \Gamma(S \otimes L^p \otimes E)$:
\begin{align}
    (U_l\psi)(x^1,x^2)
    :=&e^{ -\frac{\hbar_pl^2}{4}}\int \omega(y) {\rm P}\exp(i\int_\gamma A)B_p^{(L)}(x^1,x^2|y^1,y^2)e^{\frac{l(x^1 - y^1)}{2}}\psi(y^1,y^2), \nonumber \\
    (V_l\psi)(x^1,x^2)
    :=& e^{ -\frac{\hbar_pl^2}{4}}\int \omega(y) {\rm P}\exp(i\int_\gamma A)B_p^{(L)}(x^1,x^2|y^1,y^2)e^{\frac{l(x^2 - y^2)}{2}}\psi(y^1,y^2) .
    \label{def of ul and vl}
\end{align}
Here, $l$ is a real parameter, $\gamma$ is the shortest straight line\footnote{On the torus, the shortest 
straight line is not unique for some long paths. However, since
the Bergman kernel decays very quickly for such paths and hence 
it does not cause any problem. To define the operators more regorously,
one can insert a function that is equal to 1 on a compact support around 
the point $y$ and is vanishing elsewhere.} from $y$ to $x$
and ${\rm P}$ denotes the path ordering.
$B_p^{(L)}(x|y)$ is what is called the Bergman kernel and is defined by
\begin{align}
    B_p^{(L)}(x|y) = \sum_l \eta_l(y)\eta^{\dagger}_l(x),
    \label{bergman kernel for l}
\end{align}
where $\{ \eta_l  | l=1,2,\cdots, p\}$ is an orthonormal basis of Dirac
zero modes of $\Gamma(S \otimes L^p)$ satisfying
\begin{align}
    i\gamma^\alpha(\partial_\alpha - ipA^{(L)}_\alpha)\eta_l(x) = 0.
\end{align}
The above operators $U_l$ and $V_l$ gives linear maps on
 $\Gamma(S \otimes L^p \otimes E)$, and thus, we can consider their 
 quantization just as described in section
 \ref{Berezin-Toeplitz quantization for vector bundles}.

In order to study some properties of 
this quantization, the following asymptotic expansion
of the Bergman kernel $B_p^{(L)}(x|y)$ \cite{Ma:2010vhd} is very useful:
\begin{align}
    B_p^{(L)}(x|y) =
    \sum_{r=0}^{\infty} 
    \mathcal{P}\qty(\frac{x}{\sqrt{2\pi \hbar_p}},\frac{y}{\sqrt{2\pi\hbar_p}})
    {J}_r\qty(\frac{x}{\sqrt{2\pi \hbar_p}},\frac{y}{\sqrt{2\pi\hbar_p}}) (2\pi\hbar_p)^{\frac{r}{2}-1}P_x,
\end{align}
where $P_x$ is the projection onto the positive chirality 
and the functions $\mathcal{P}$ and $J_r$
are given as
\begin{align}
    &\mathcal{P}(x,y) = \exp(-\frac{\pi}{2}(|x^1-y^1|^2+|x^2-y^2|^2-2i(x^1y^2-y^1x^2))), \nonumber\\
    &J_0(x,y)=1,\quad J_1(x,y) = 0, \quad \cdots.
\end{align}
When $p$ is sufficiently large (or equivalently when $\hbar_p$ is small), we can calculate as
\begin{align}
    &\mathcal{P}\qty(\frac{x}{\sqrt{2\pi \hbar_p}},\frac{y}{\sqrt{2\pi\hbar_p}})e^{-\frac{\epsilon(x^1 - y^1)}{2}}e^{-\frac{\hbar_p\epsilon^2}{4}} (2\pi\hbar_p)^{\frac{r}{2}-1}\nonumber\\
    &= \frac{1}{4\pi\hbar_p}\exp(-\frac{1}{4\hbar_p}(|x^1-y^1+\hbar_p\epsilon|^2+|x^2-y^2|^2))
    \cdot 2(2\pi\hbar_p)^{\frac{r}{2}}\exp(\frac{i}{2\hbar_p}(x^1y^2-y^1x^2)) 
    \nonumber\\
    &\sim  \delta(x^a-y^a+\delta^{a1}\hbar_p\epsilon)\cdot 2(2\pi\hbar_p)^{\frac{r}{2}}\exp(-\frac{i\epsilon x^2}{2})
    \label{limit of bergman kernel}
\end{align}
Thus, the operators $U_l$ and $V_l$ are 
 essentially the Wilson lines along $x^1$ and $x^2$ directions, respectively
 in the large-$p$ limit.

We consider the quantization of $U_l$ and $V_l$. As shown in
 appendix~\ref{Some properties of quantized Wilson lines},
the Toeplitz operators of $U_l$ and $V_l$ satisfy 
\begin{align}
    \label{eq:pro_her}
    T_p(U_l)^{\dagger} = T_p(U_{l}^{\dagger}),\quad
    T_p(V_l)^{\dagger} = T_p(V_{l}^{\dagger}),
\end{align}
and also
\begin{align}
    \label{eq:pro_asy}
    T_p(U_{l_1})T_p(V_{l_2}) = T_p(U_{l_1}V_{l_2}) + O(\hbar_p).
\end{align}

Let us then consider the quantization of a single plaquette,
\begin{align}
    \label{eq:large_N}
    &{\rm Re}{\rm Tr}T_p(e^{-i\hbar_p\epsilon^2}U_{-\epsilon}V_{-\epsilon}U_{\epsilon}V_{\epsilon}) \nonumber\\
    &= e^{-i\hbar_p\epsilon^2}\qty(e^{ -\frac{\hbar_p\epsilon^2}{4}})^4{\rm Re}\int \omega' \sum_I {\rm tr}
    \left[
    \psi^{\dagger}_I(x){\rm P}\exp(i\int_{\gamma'(x|v)} A)\right.
    \nonumber\\
    &\left.
    B_p^{(L)}(x|y)
    B_p^{(L)}(y|z)
    B_p^{(L)}(z|w)
    B_p^{(L)}(w|v)
    e^{-\frac{\epsilon(x^1 - y^1)}{2}}
    e^{-\frac{\epsilon(x^2 - y^2)}{2}}
    e^{\frac{\epsilon(x^1 - y^1)}{2}}
    e^{\frac{\epsilon(x^2 - y^2)}{2}}\psi_I(v)
    \right].
\end{align}
Here, we used the shorthand notation for the integration measure, 
\begin{align}
   \int \omega' &= \int \omega(x) \int \omega(y) \int \omega(z)\int \omega(w)\int \omega(v) .
\end{align}
and the total path $\gamma'(x|v)$ consists of the straight lines from $v$ to $w$, $w$ to $z$, $z$ to $y$, and $y$ to $x$. By using 
the relation (\ref{limit of bergman kernel}) in the large-$p$ limit, 
we can perform four integration.
Furthermore, we can use the formula \cite{Ma:2010vhd} for the diagonal 
asymptotic expansion of the Bergman kernel, given by
\begin{align}
     \sum_I \psi_I(x)\psi^{\dagger}_I(x) 
    = \sum_r b^{(E)}_r(x)(2\pi \hbar_p)^{r-1} P_x ,
    \label{diagonal aym exp}
\end{align}
where the coefficient $b^{(E)}_r(x)$ are given as
\begin{align}
    b^{(E)}_0(x) = {\rm Id}_E,
    \quad b^{(E)}_1(x) = \frac{1}{4\pi}F^{(E)}_{12},
    \quad \cdots .
\end{align}
By using (\ref{diagonal aym exp}), we obtain
\begin{align}
    &{\rm Re}{\rm Tr}T_p(e^{-i\hbar_p\epsilon^2}U_{-\epsilon}V_{-\epsilon}U_{\epsilon}V_{\epsilon})
    = \frac{8}{\pi\hbar_p} \int \omega(x) {\rm tr}
    \qty[
    C(x) - \epsilon^4\hbar_p^4\qty(F_{12}(x))^2 + O(\hbar_p^{9/2})
    ],
\end{align}
where $F$ is the curvature of $A$ and the function $C$ is defined by
\begin{align}
    C(x) = \sum_{r\leq 4}b_{r_1}(x)&
    J_{r_2}\qty(x/\sqrt{2\pi \hbar_p},y/\sqrt{2\pi\hbar_p})
    J_{r_3}\qty(y/\sqrt{2\pi \hbar_p},z/\sqrt{2\pi\hbar_p}) \nonumber\\
    &\times 
    J_{r_4}\qty(z/\sqrt{2\pi \hbar_p},w/\sqrt{2\pi\hbar_p})
    J_{r_5}\qty(w/\sqrt{2\pi \hbar_p},x/\sqrt{2\pi\hbar_p})
    (2\pi\hbar_p)^r
\end{align}
with $r={r_1}+\frac{{r_2}+{r_3}+{r_4}+{r_5}}{2}$.
Finally, by using (\ref{eq:pro_asy}), we find  that in the large-$p$ limit,
\begin{align}
    \frac{\pi e^{-i\hbar_p\epsilon^2}}{8g^2\hbar_p^3\epsilon^4}{\rm Re}{\rm Tr}\qty(T_p(U_{\epsilon})^{\dagger}T_p(V_{\epsilon})^{\dagger}T_p(U_{\epsilon})T_p(V_{\epsilon}))
    \sim -\frac{1}{4g^2}\int \omega {\rm tr}(F_{12})^2,
\end{align}
where we ignored the field independent term on the right-hand side.
The right-hand side is just the standard action for the pure Yang-Mills theory
while the left-hand side is the action of the Eguchi-Kawai model.
The only subtlety is the unitarity of the Toeplitz operators. 
However, because of relations such as $U_{l}^\dagger U_{l}= 1+O(\hbar_p)$ 
and the asymptotic 
properties of $T_p$, they become unitary matrices in $\hbar_p \rightarrow 0$. 
Thus, we find that the model is obtained from the matrix regularization 
of Wilson line operators accompanied with the Bergman kernel.

This correspondence holds at least at the classical level.
At the quantum mechanical level, there are potentially subtle problems in 
the large-$N$ reduction such as the $U(1)^D$ symmetry breaking 
\cite{Bhanot:1982sh}
and the UV/IR mixing \cite{Douglas:2001ba}. 
Because of these problems, 
the above argument does not necessarily 
garantee the corrspondence at the quantum level.
In order to study the quantum correspondence,
it would be important to translate the action of
Eguchi-Kawai model into the continuum form including the 
$\hbar_p$ correction to see whether the correction produces 
those problems at the quantum level or not.

The definition (\ref{def of ul and vl}) may look awkward. 
It may be more useful to express the Bergman kernel in terms of the 
path integral \cite{Douglas:2009fvp} and define $U_l$ and $V_l$ as 
a path integral of Wilson lines.
Such expression may be useful for generalizing above discussion to
the case of curved nontrivial manifold.

%%%%%%%%%%%%%%%%%%%%%%%%%%%%%%%%%%%%%%%%%%%%%%%%%%%%%%%%%%%%%
%%%%%%%%%%%%%%%%%%%%%%%%%%%%%%%%%%%%%%%%%%%%%%%%%%%%%%%%%%%%%
\section{Summary and discussion}
\hspace{0.51cm}
%%%%%%%%%%%%%%%%%%%%%%%%%%%%%%%%%%%%%%%%%%%%%%%%%%%%%%%%%%%%%
%%%%%%%%%%%%%%%%%%%%%%%%%%%%%%%%%%%%%%%%%%%%%%%%%%%%%%%%%%%%%
In this paper, we studied how gauge fields can be described in 
matrix models. We considered the Berezin-Toeplitz quantization, which is a 
concrete realization of the matrix regularization, as a tool 
of connecting the matrices and continuous gauge fields.
At first sight, since such quantization applies only to matter fields, 
there seems no room for gauge fields. 
However, when the quantization is applied to embedding functions 
of brane-like objects, tangential and transverse fluctuations of the embedding 
functions certainly contains the degrees of freedom for the gauge fields 
and scalar fields, respectively.
We defined the Seiberg-Witten map that connects finite-size matrices and those fields 
and obtained an explicit form of the map (\ref{S2SWmap}) for $S^2$ up to the 
next leading order in $1/N$.
The emergence of the local gauge symmetry from matrices is quite nontrivial, 
and we argued that this originates from the gauge redundancy (\ref{gauge symmetry of Teoplitz operator}) in constructing the quantization map.

We also studied relationships with other formulations of gauge fields.
The first formulation is the one discussed in the context of tachyon condensation \cite{Terashima:2005ic}.
We argued that the gauge field defined in this paper is also related to the 
gauge field in tachyon condensation, through a redefinition of the field variables.
The second formulation we discussed is Eguchi-Kawai model.
We showed that variables in this model can be interpreted as 
quantized Wilson line operators in our setup.

It is important to consider the inverse problem of our problem
to understand more deeply the relation between matrices and gauge fields.
The inverse problem of fuzzy geometry (namely, finding a classical geometry from 
quantized geometry) has been discussed in
\cite{Berenstein:2012ts,Ishiki:2015saa,Schneiderbauer:2016wub,Ishiki:2016yjp,Sako:2022pid} and it would be interesting to consider the dual description 
of our findings (A possible candidate for such description is 
provided by the Berry connection \cite{Ishiki:2016yjp} in the matrix geometry.).
Such description will be useful in understanding physics
of tachyon condensation \cite{Asakawa:2018gxf,Terashima:2018tyi}.

%%%%%%%%%%%%%%%%%%%%%%%%%%%%%%%%%%%%%%%%%%%%%%%%%%%%%%%%%%%%%
\section*{Acknowledgments}
%%%%%%%%%%%%%%%%%%%%%%%%%%%%%%%%%%%%%%%%%%%%%%%%%%%%%%%%%%%%%
We thank Seiji Terashima for useful discussion.
The work of H. A. and G. I. was supported  
by JSPS KAKENHI (Grant Numbers 21J12131 and 23K03405, respectively). 
This work of S. K. was supported by JST, the establishment of university fellowships towards the creation of science technology innovation, Grant Number JPMJFS2106.

%%%%%%%%%%%%%%%%%%%%%%%%%%%%%%%%%%%%%%%%%%%%%%%%%%%%
\begin{appendix}
\numberwithin{equation}{section}
\setcounter{equation}{0}

%%%%%%%%%%%%%%%%%%%%%%%%%%%%%%%%%%%%%%%%%%%%%%%%%%%%%%%%%%%%%
%%%%%%%%%%%%%%%%%%%%%%%%%%%%%%%%%%%%%%%%%%%%%%%%%%%%%%%%%%%%%
\section{Gauge symmetry of the Toeplitz operator}
\label{The gauge symmetry of the Toeplitz operator}
%%%%%%%%%%%%%%%%%%%%%%%%%%%%%%%%%%%%%%%%%%%%%%%%%%%%%%%%%%%%%
%%%%%%%%%%%%%%%%%%%%%%%%%%%%%%%%%%%%%%%%%%%%%%%%%%%%%%%%%%%%%
In this appendix, we show that the Toeplitz operator (\ref{hatX})
is invariant under the transformation (\ref{gauge symmetry of Teoplitz operator}).
In this argument, we should be careful about the 
$A^{(E)}$-dependence of $\tilde{X}^A$. 
The functional form of $\tilde{X}^A(B, \chi)$ is determined by finding the 
SW map, which is defined from the asymptotic properties of Toeplitz operators. 
Since the definition of Toeplitz operators depends on $A^{(E)}$, $\tilde{X}^A$
 also implicitly depends on $A^{(E)}$. Hence, it is appropriate to write
 $\tilde{X}^A(A^{(E)}|B, \chi)$, where we omit the $X$-dependence here.  
With this notation, the invariance of the Toeplitz operator is 
 explicitly expressed as 
 \als{
    \label{eq:appA_goal} 
    T_p(X^A 1_k + \hbar_p \tilde{X}^A(A^{(E)}| B, \chi) )
    =T_p^{(U)}(X^A 1_k +\hbar_p \tilde{X}^{A}(A^{U(E)}| B^U, \chi^U)),
}
where $A^{U(E)}$, $B^U$ and $\chi^U$ are defined in
(\ref{gauge symmetry of Teoplitz operator}) and 
$T_p^{(U)}$ is the Toeplitz operator made from $A^{U(E)}$.

There are several useful properties 
of the Toeplitz operator for deriving \eqref{eq:appA_goal}.
Under gauge transformation of $A^{(E)}$, the Dirac operator transforms 
covariantly as $D^{U(E)}=UD^{(E)}U^{-1}$.
Then, the zero modes of $D^{U(E)}$ can be chosen as $\psi^U_I=U\psi_I$, where $\psi_I$ are the zero modes of $D^{(E)}$. Thus, there is a relation,
\als{
    \label{eq:UphiU}
    T_p^{(U)}(\varphi) = T_p(U^{-1}\varphi U)
}
for any $\varphi \in \Gamma({\rm Hom}(E,E))$.
The asymptotic expansion of $T_p^{(U)}$ is written as
\als{
    T^{(U)} (\varphi')T^{(U)} (\varphi) = \sum_{s=0}^{\infty} \hbar_{p}^s T^{(U)} (C^{(U)}_s(\varphi',\varphi)).
}
From the covariance of $C_s$, we also have
\als{
    C^{(U)}_s(\varphi',\varphi) = U C_s(U^{-1}\varphi'U,U^{-1}\varphi U) U^{-1}.
    \label{eq:asym_gauge}
}

Let us prove \eqref{eq:appA_goal}. By putting $\varphi = X^A 1_k +
\hbar_p U \tilde{X} U^{-1}$ in \eqref{eq:UphiU}, we have 
\als{
    T_p(X^A 1_k + \hbar_p \tilde{X}^A(A^{(E)}| B, \chi) )
    =T_p^{(U)}(X^A 1_k +\hbar_p U \tilde{X}^A(A^{(E)}| B, \chi) U^{-1}) .
} 
Below, we will show that $U \tilde{X}^A(A^{(E)}| B, \chi)U^{-1}= 
\tilde{X}^A(A^{U(E)}| B^U, \chi^U)$, which implies \eqref{eq:appA_goal}.
More specifically,  if we express $U \tilde{X}^A U^{-1}$ 
as a linear combination of $UA_aU^{-1}$ and $U\phi_iU^{-1}$ 
by using (\ref{expantion of fluctuations}), 
we will show that they satisfy
\begin{align}
&UA_a(A^{(E)}| B, \chi)U^{-1} = A_a(A^{U(E)}| B^U, \chi^U),  \nonumber\\
&U\phi_i (A^{(E)}| B, \chi)U^{-1} = \phi_i (A^{U(E)}| B^U, \chi^U).
\label{invariance for a and phi}
\end{align}
Let us first consider $UA_aU^{-1}$. From the SW map (\ref{SW expansion}), we have an expansion, 
\begin{align}
    \Delta_{a}(A^{U(E)}|B^U,\chi^U)& := U A_a(A^{(E)}|B,\chi)U^{-1}, \nonumber\\
    &=B^U_a-A^{U(E)}_a +\hbar_p U B_{1,a}(A^{(E)}|B,\chi)U^{-1} +\cdots. 
    \label{def of Deltaa}
\end{align}
Notice that $\Delta_{a}(A^{U(E)}|B^U,\chi^U)$ satisfies the following relation:
\als{
    \Delta_{a}(A^{U(E)}|B^U,\chi^U)+\delta^{(U)}_{U\alpha U^{-1}}\Delta_{a}(A^{U(E)}|B^U,\chi^U)
    &=U A_a(A^{(E)}|B,\chi)U^{-1} + U \delta_{\alpha}A_a(A^{(E)}|B,\chi)U^{-1}\\
    &=U A_a(A^{(E)}|B+\delta_{\beta}B,\chi+\delta_{\beta}\chi) U^{-1}\\
    &=\Delta_{a}(A^{U(E)}|B^U+\delta_{U\beta U^{-1}}B^U,\chi^U+\delta_{U\beta U^{-1}}\chi),
    \label{SW for delta}
}
where, $\delta^{(U)}_{U\alpha U^{-1}}$ stands for 
the gauge transformation under the background field $A^{U(E)}$ defined by
\als{
    \delta^{(U)}_{U\alpha U^{-1}} \Delta_a&=i\sum_{s=1}^{\infty}\hbar_p^{s-1}W_{ab}g^{bc}\nabla_c X^A(D^{(U)}_s(U\alpha U^{-1},X^A1_k)+D^{(U)}_{s-1}(U\alpha U^{-1},U\Tilde{X}^{A}U^{-1})).
}
The first equality of (\ref{SW for delta}) is obtained from
\als{
    \delta^{(U)}_{U\alpha U^{-1}} \Delta_a
    =U\delta_{\alpha}A_a U^{-1},
}
which follows from \eqref{eq:asym_gauge}. The second equality
(\ref{SW for delta}) comes from the condition \eqref{SWeq} of the SW map.
The third equality of (\ref{SW for delta}) follows from
\als{
    B^U_a+\delta_{U\beta U^{-1}}B^U_a
    &= (B_a +\delta_\beta B_a)^U, \\
    \chi^U_i+\delta_{U\beta U^{-1}}\chi^U_i
    &= (\chi_i +\delta_\beta \chi_i)^U.
}
The relation (\ref{SW for delta}) is nothing but the compatibility condition for 
the SW map. Furthermore, from (\ref{def of Deltaa}), the lowest order term 
of $\Delta_a$ is given by $B^U_a-A^{U(E)}_a$. These are just the defining 
equations for the function $A(A^{U(E)}|B^U,\chi^U)$. Thus, we find that 
$\Delta(A^{U(E)}|B^U,\chi^U)=A(A^{U(E)}|B^U,\chi^U)$, which is equivalent to
the first equation of (\ref{invariance for a and phi}).
The second equation of (\ref{invariance for a and phi}) can be shown in 
a similar way.

%%%%%%%%%%%%%%%%%%%%%%%%%%%%%%%%%%%%%%%%%%%%%%%%%%%%%%%%%%%%%
%%%%%%%%%%%%%%%%%%%%%%%%%%%%%%%%%%%%%%%%%%%%%%%%%%%%%%%%%%%%%
\section{The $\pm$ notation}
\label{The plus-minus notation}
%%%%%%%%%%%%%%%%%%%%%%%%%%%%%%%%%%%%%%%%%%%%%%%%%%%%%%%%%%%%%
%%%%%%%%%%%%%%%%%%%%%%%%%%%%%%%%%%%%%%%%%%%%%%%%%%%%%%%%%%%%%
For any vector $A_a (a=1,2)$ in the orthonormal frame, we 
take the combination $A_{\pm}=\frac{1}{\sqrt{2}}(A_1\pm iA_2)$.
In this appendix, we describe this notation.

The Poisson tensor $W^{ab}=e^{a}_\alpha e^b_\beta W^{\alpha \beta}$ in 
the orthonormal coordinate has components $W^{12}=-W^{21}=1$
and thus for any vectors $A$ and $B$, we have
\als{
    \begin{split}
        W^{ab}A_a B_b &= A_1 B_2 - A_2 B_1 = i(A_+ B_- - A_- B_+).
    \end{split}
}
This shows $W^{+-}=i$.

Next, let us consider the action of commutator $[\nabla_- , \nabla_+]$ on $S^2$.
For any cotangent vector field $A_c=e_c^\alpha A_\alpha$ in the orthonormal 
frame, we have
\als{
    [\nabla_a , \nabla_b]A_c=R^d_{cab}A_d.
}
Here, $R^d_{cab}$ is Riemann tensor and $R^1_{212}=\frac{R}{2}=1$ on $S^2$ with the radius $1$.
Then, we obtain the following equation
\als{
    [\nabla_- , \nabla_+]A_+=\frac{R}{2}A_+=A_+
}
Similarly, we also have $[\nabla_- , \nabla_+]A_-=A_-$.

%%%%%%%%%%%%%%%%%%%%%%%%%%%%%%%%%%%%%%%%%%%%%%%%%%%%%%%%%%%%%
%%%%%%%%%%%%%%%%%%%%%%%%%%%%%%%%%%%%%%%%%%%%%%%%%%%%%%%%%%%%%
\section{Formulas for isometric embedding of $S^2$}
\label{A formula for the isometric embedding of S2}
%%%%%%%%%%%%%%%%%%%%%%%%%%%%%%%%%%%%%%%%%%%%%%%%%%%%%%%%%%%%%
%%%%%%%%%%%%%%%%%%%%%%%%%%%%%%%%%%%%%%%%%%%%%%%%%%%%%%%%%%%%%
In this Appendix, we derive a useful equation that the 
embedding function of $S^2$ satisfies.

Let us consider the isometric embedding $X^A:S^2\rightarrow \mathbb{R}^3$ 
satisfying (\ref{xx=1}) and
\als{
    \label{eq:XX}
    g_{\alpha \beta}=\partial_\alpha X^A\partial_\beta X^A,
}
where $g$ is the metric on $S^2$.
The tangential vectors $\partial_\alpha X^A $ and the radial 
vector $X^A$ form a basis of the three dimensional vector space 
and other vectors such as $\nabla_\alpha \partial_\beta X^A$ should be able 
to be expanded in terms of the basis.
Below, we will show 
\als{
    \label{eq:XnabX}
    \nabla_\alpha \partial_\beta X^A=-g_{\alpha \beta }X^A.
}
First, note that the Christoffel symbol 
for the metric $g$ is given in terms of $X^A$ as
\als{
    \Gamma^{\gamma}_{\alpha \beta}&=\frac{1}{2}g^{\gamma \delta}
    (\partial_\alpha g_{\delta \beta}+\partial_\beta g_{\delta \alpha}-\partial_\delta g_{\alpha \beta})
    =\partial^\gamma X^A\partial_\alpha \partial_\beta X^A.
}
From this expression, we find that
\als{
    (\partial^\alpha X^A)(\nabla_\beta \partial_\gamma X^A)
    =\Gamma^\alpha_{\beta \gamma }-\Gamma^\delta_{\beta \gamma }\delta_{\delta}^\alpha=0.
}
Thus, $\nabla_\beta \partial_\gamma X^A$ is perpendicular to the sphere and thus 
should be proportional to $X^A$.
The factor of proportionality is deduced from
\als{
    X^A\nabla_\alpha \partial_\beta X^A&=\nabla_\alpha (X^A\partial_\beta X^A)-\partial_\alpha X^A\partial_\beta X^A 
    = -g_{\alpha \beta },
}
where, we used $X^A\partial_\alpha X^A=0$ which follows from (\ref{xx=1}).
This shows the equation (\ref{eq:XnabX}).

%%%%%%%%%%%%%%%%%%%%%%%%%%%%%%%%%%%%%%%%%%%%%%%%%%%%%%%%%%%%%
%%%%%%%%%%%%%%%%%%%%%%%%%%%%%%%%%%%%%%%%%%%%%%%%%%%%%%%%%%%%%
\section{Derivation of SW map on $S^2$}
\label{The derivation of the SW map of S2}
%%%%%%%%%%%%%%%%%%%%%%%%%%%%%%%%%%%%%%%%%%%%%%%%%%%%%%%%%%%%%
%%%%%%%%%%%%%%%%%%%%%%%%%%%%%%%%%%%%%%%%%%%%%%%%%%%%%%%%%%%%%
In this Appendix, we derive the SW map \eqref{S2SWmap} on $S^2$.

We first calculate $\chi_1$.
By substituting
the expansion $\phi = \chi + \hbar_p \chi_1 + \cdots$ into the transformation law \eqref{varia_Aphi}, and picking up the term of $O(\hbar_p)$, 
we find that
\als{
    \label{SWeq1ji}
    \delta_\beta \chi_1 
    &= i[\beta, \chi_1] + i[\beta_1, \chi] - B_-\nabla_+\beta - \nabla_-\beta B_+ + i\nabla_-\chi\nabla_+\beta - i\nabla_-\beta\nabla_+\chi \\
    & = i[\beta, \chi_1] + i[\beta_1, \chi] - B_-(\delta_\beta B_{+} - i[\beta,B_{+}]) - (\delta_\beta B_{-} - i[\beta,B_{-}]) B_+ \\
    &\quad\quad\quad\quad\quad\quad\quad\quad\quad+ i\nabla_-\chi(\delta_\beta B_{+} - i[\beta,B_{+}]) - i(\delta_\beta B_{-} - i[\beta,B_{-}])\nabla_+\chi\\
    & = \delta_\beta \qty(-B_-B_+ + i\nabla_- \chi B_+ - iB_-\nabla_+ \chi -\frac{1}{2}B_-[B_+,\chi] + \frac{1}{2}[B_-,\chi]B_+)\\
    &\quad\quad+ i\qty[\beta, \chi_1 + B_-B_+ - i\nabla_- \chi B_+ + iB_-\nabla_+ \chi +\frac{1}{2}B_-[B_+,\chi] - \frac{1}{2}[B_-,\chi]B_+]\\
    &\quad\quad\quad\quad\quad\quad\quad\quad\quad\quad\quad\quad\quad\quad\quad\quad\quad\quad+ i\qty[\beta_1 -\frac{i}{2}\nabla_-\beta B_+ + \frac{i}{2}B_-\nabla_+ \beta,\chi] ,
}
where in the second equality, we used $\nabla_{\pm}\beta = \delta_\beta B_{\pm} - i[\beta,B_{\pm}]$. The last expression is integrable and 
 we obtain the solution $\chi_1$ and $\beta_1$ as in (\ref{S2SWmap}).

We can calculate $B_1$ in the similar manner.
We first consider the minus component $B_{1-}$.
Its transformation law is derived  from  \eqref{varia_Aphi}, 
and is calculated as
\als{
    \label{SWeq1jiB}
    \delta_\beta B_{1-} 
    &= \nabla_-\beta_1 + i[\beta, B_{1-}] + i[\beta_1, B_-] - \frac{R}{4}\nabla_- \beta + \nabla_-\beta \chi + i\nabla_-B_-\nabla_+\beta - i\nabla_-\beta\nabla_+ B_- \\
    &= \nabla_-\qty(-\frac{i}{2}B_-\nabla_+\beta + \frac{i}{2}\nabla_-\beta B_+) + i\qty[-\frac{i}{2}B_-\nabla_+\beta + \frac{i}{2}\nabla_-\beta B_+, B_-] \\
    &\quad\quad\quad\quad\quad\quad
    + i[\beta, B_{1-}] - \frac{R}{4}\nabla_- \beta + \nabla_-\beta \chi + i\nabla_-B_-\nabla_+\beta - i\nabla_-\beta\nabla_+ B_- \\
    &= \nabla_-\qty(-\frac{i}{2}B_-(\delta_\beta B_{\pm} - i[\beta,B_{\pm}]) + \frac{i}{2}(\delta_\beta B_{\pm} - i[\beta,B_{\pm}]) B_+) \\
    &\quad\quad\quad\quad
    + i\qty[-\frac{i}{2}B_-(\delta_\beta B_{\pm} - i[\beta,B_{\pm}]) + \frac{i}{2}(\delta_\beta B_{\pm} - i[\beta,B_{\pm}]) B_+, B_-] \\
    &\quad\quad\quad\quad\quad
    + i[\beta, B_{1-}] - \frac{R}{4}(\delta_\beta B_{\pm} - i[\beta,B_{\pm}])+ (\delta_\beta B_{\pm} - i[\beta,B_{\pm}]) \chi \\
    &\quad\quad\quad\quad\quad\quad
    + i\nabla_-B_-(\delta_\beta B_{\pm} - i[\beta,B_{\pm}])- i(\delta_\beta B_{\pm} - i[\beta,B_{\pm}])\nabla_+ B_- \\
    &= \delta_\beta\qty(-\frac{R}{4}B_- + B_-\chi +\frac{i}{2}(\nabla_-B_-)B_+ - iB_-\nabla_+B_- + \frac{i}{2}B_-\nabla_-B_+ + \frac{1}{2}B_-[B_-,B_+])\\
    &\quad\quad\quad
    +i\qty[\beta, - B_{1,-} -\frac{R}{4}B_- + B_-\chi +\frac{i}{2}(\nabla_-B_-)B_+ - iB_-\nabla_+B_- + \frac{i}{2}B_-\nabla_-B_+ + \frac{1}{2}B_-[B_-,B_+]],
}
where in the second equality, we substituted the expression of $\beta_1$ in (\ref{S2SWmap}) and 
in the third equality, we used $\nabla_{\pm}\beta = \delta_\beta B_{\pm} - i[\beta,B_{\pm}]$.
By integrating the above expression, we obtain the solution for 
$B_{1-}$ as shown in (\ref{S2SWmap}). 
$B_{1+}$ can easily be found from $(B_{1-})^{\dagger}=B_{1+}$.

Note that the higher order terms of the SW map have ambiguities in 
adding gauge covariant quantities. We have considered the simplest case 
where such ambiguous terms are simply put zero.

%%%%%%%%%%%%%%%%%%%%%%%%%%%%%%%%%%%%%%%%%%%%%%%%%%%%%%%%%%%%%
%%%%%%%%%%%%%%%%%%%%%%%%%%%%%%%%%%%%%%%%%%%%%%%%%%%%%%%%%%%%%
\section{Calculation for the qubic matrix model}
\label{The calculation for the qubic matrix model}
%%%%%%%%%%%%%%%%%%%%%%%%%%%%%%%%%%%%%%%%%%%%%%%%%%%%%%%%%%%%%
%%%%%%%%%%%%%%%%%%%%%%%%%%%%%%%%%%%%%%%%%%%%%%%%%%%%%%%%%%%%%
In this appendix, we show a derivation of 
(\ref{BF plus quantum corrections}) in detail.

\subsection{Preliminary}
Here, we show some relations that are needed for the derivation of 
(\ref{BF plus quantum corrections}) including the $1/p$ corrections.

We first calculate $\int \omega {\rm tr} \chi_2$.
The gauge transformation law for $\chi_2$ is obtained from
\eqref{varia_Aphi} as
\begin{align}
    \delta_\beta \chi_2
    = & i[\beta, \chi_2] + i[\beta_2, \chi] + i[\beta_1,\chi_1] \nonumber\\
    & - B_-\nabla_+ \beta_1 - B_{1-}\nabla_+ \beta - \nabla_- \beta_1 B_+ - \nabla_- \beta B_{1+}\nonumber\\
    & + i\nabla_-\chi_1 \nabla_+\beta + i\nabla_-\chi \nabla_+\beta_1 - i\nabla_-\beta\nabla_+\chi_1  - i\nabla_-\beta_1\nabla_+\chi \nonumber\\
    & -\frac{R}{4}\qty(- B_-\nabla_+ \beta - \nabla_- \beta B_+ + i\nabla_-\chi \nabla_+\beta - i\nabla_-\beta\nabla_+\chi) \nonumber\\
    & + \nabla_- B_-\nabla^2_+\beta + \nabla^2_-\beta\nabla_+ B_+
    + \frac{i}{2}\nabla^2_-\beta\nabla^2_+\chi - \frac{i}{2}\nabla^2_-\chi\nabla^2_+\beta.
\end{align}
By taking the trace and integration, we obtain
\begin{align}
    \int \omega {\rm tr} \delta_\beta\chi_2 
    &= \int \omega {\rm tr} \left(
    - B_-\nabla_+ \beta_1 - B_{1-}\nabla_+ \beta - \nabla_- \beta_1 B_+ - \nabla_- \beta B_{1+} \right. \nonumber\\
    &\left. \quad\quad\quad\quad\quad\quad\quad\quad\quad
    -\frac{R}{4}\qty(- B_-\nabla_+ \beta - \nabla_- \beta B_+ )+ \nabla_- B_-\nabla^2_+\beta + \nabla^2_-\beta\nabla_+ B_+
    \right) \nonumber\\
    &= \int \omega {\rm tr} \left(
    - B_-\nabla_+\qty(-\frac{i}{2}B_-\nabla_+\beta + \frac{i}{2}\nabla_-\beta B_+)
    - \nabla_-\qty(-\frac{i}{2}B_-\nabla_+\beta + \frac{i}{2}\nabla_-\beta B_+) B_+
    \right. \nonumber\\
    &\left. \quad\quad\quad\quad\quad
    - B_{1-}\nabla_+ \beta
    - \nabla_- \beta B_{1+} 
    -\frac{R}{4}\qty(- B_-\nabla_+ \beta - \nabla_- \beta B_+ )
    + \nabla_- B_-\nabla^2_+\beta
    + \nabla^2_-\beta\nabla_+ B_+
    \right) \nonumber\\
    &= \int \omega {\rm tr} \left(
    \delta_\beta \qty(-B_{1-}B_+ - B_-B_{1+} + B_-\chi B_+ +\nabla_-B_-\nabla_+B_+)
    \right).
    \label{delta trace chi 2}
\end{align}
Here, in obtaining the first equality, we used the equation,
\als{
    \int \omega {\rm tr}\qty(\frac{i}{2}\nabla^2_-\beta\nabla^2_+\chi - \frac{i}{2}\nabla^2_-\chi\nabla^2_+\beta) =0,
}
which is a consequence of the relation $[\nabla_+, \nabla_-]T_{+-}=
-\frac{R}{2}T_{+-}+ \frac{R}{2}T_{+-}=0$ for any rank 2 tensor $T$.
In obtaining the second equality, we substituted $\beta_1$ in (\ref{S2SWmap}).
By integrating (\ref{delta trace chi 2}), we obtain
\als{
    \int \omega {\rm tr} \chi_2 
    &= \int \omega {\rm tr} \left(
    \qty(-B_{1-}B_+ - B_-B_{1+} + B_-\chi B_+ +\nabla_-B_-\nabla_+B_+) + Q(B,\chi)
    \right),
}
where $Q(B,\chi)$ is a function of $B$ and $\chi$ satisfying
\als{
    \int \omega {\rm tr} \qty(\delta_\beta Q(B,\chi)) = 0.
}
Note that $\chi_2$ originally has an ambiguity in adding a gauge covariant 
quantity. The above function $Q$ represents this ambiguity and we will consider
the case of $Q=0$ for simplicity. We then obtain
\als{
    \int \omega {\rm tr} \chi_2 
    &= \int \omega {\rm tr} \left(
    \qty(- B_-B_+ -B_{1-}B_+ - B_-B_{1+} + B_-\chi B_+ +\nabla_+B_-\nabla_-B_+)
    \right).
    \label{eq:core_chi2}
}

From (\ref{S2SWmap}), we can easily find that $\int \omega {\rm tr} \chi_1$,
$\int \omega {\rm tr} \chi \chi_1$ and
$\int \omega {\rm tr}(iF_{12}^0 \chi_1)$, 
 are given as
\als{
    \label{eq:core_chi1}
    \int \omega {\rm tr} \chi_1
    &= \int \omega {\rm tr}\qty(-B_-B_+ + F^0_{12}\chi),
}
\als{
    \label{eq:core_chichi1}
    \int \omega {\rm tr} \chi \chi_1
    &= \int \omega {\rm tr}\qty(-\chi B_-B_+ + \frac{i}{2}\nabla_-\chi[B_+,\chi] + \frac{i}{2}\nabla_+\chi[B_-,\chi] + \frac{1}{2}\chi^2 F^0_{12} + \frac{1}{2}[B_-, \chi][B_+, \chi]),
}
and
\als{
    \label{eq:core_Fchi}
    &\int \omega {\rm tr}(iF_{12}^0 \chi_1)
   = \int \omega {\rm tr} \left(
    -\nabla_-B_+ B_- B_+ + \nabla_+B_- B_- B_+ + \frac{i}{2}[B_-,B_+]^2 \right. \\
    &\quad\quad\quad\quad\quad + i B_+ \nabla_- B_+ \nabla_- \chi - i B_+ \nabla_+ B_- \nabla_- \chi - i  \nabla_- B_+ B_- \nabla_+ \chi + i \nabla_+ B_- B_- \nabla_+ \chi \\
    &\quad\quad\quad\quad\quad - \frac{1}{2} \nabla_- B_+ B_- [B_+, \chi] + \frac{1}{2} \nabla_- B_+ [B_-, \chi] B_+ + \frac{1}{2} \nabla_+ B_- B_- [B_+, \chi] - \frac{1}{2} \nabla_+ B_- [B_-, \chi] B_+ \\
    &\left.\quad\quad\quad\quad\quad + [B_-, B_+] \nabla_-\chi B_+ -  [B_-, B_+] B_- \nabla_+\chi
    + \frac{i}{2}[B_-,B_+]B_- [B_+,\chi] - \frac{i}{2}[B_-,B_+][B_-,\chi]B_+
    \right).
}
From (\ref{S2SWmap}), we also find that
\als{
    \label{eq:core_F_1chi}
    &\int \omega {\rm tr}((\nabla_-B_{1+}-\nabla_+B_{1-} -i[B_{1-},B_+] + i[B_{-},B_{1+}])\chi) \\
    &= \int \omega {\rm tr}\left(
    -\frac{R}{4}iF_{12}^0\chi + \frac{R}{4}i[B_-,B_+]\chi - i B_-B_+\nabla_-\nabla_+ \chi -\nabla_-B_+ B_- B_+ + \nabla_+B_- B_- B_+ + \frac{i}{2}[B_-,B_+]^2 \right.\\
    &\quad\quad\quad\quad\quad\quad + \frac{1}{2}[B_+,\chi]\nabla_-\chi +\frac{1}{2}[B_-, \chi]\nabla_+ \chi +\frac{1}{2}(\nabla_- B_+ - \nabla_+B_-)\chi^2 i[B_-,\chi][B_+,\chi] - i[B_-,B_+]\chi^2 \\
    &\quad\quad\quad\quad\quad\quad + F^0_{12}[B_+,\nabla_-\chi] + F^0_{12}[B_-,\nabla_+\chi] - \frac{i}{2} F_{12}^0[B_-,[B_+,\chi]] - \frac{i}{2} F_{12}^0[B_+,[B_-,\chi]]\\
    &\left. \quad\quad\quad\quad\quad\quad+\frac{1}{2}[B_+,B_-][\nabla_-B_+,\chi] + \frac{1}{2}[B_+,B_-][\nabla_+B_-,\chi] - i\nabla_+B_- B_+\nabla_-\chi - iB_- \nabla_-B_+\nabla_+\chi
    \right).
}

%%%%%%%%%%%%%%%%%%%%%%%%%%%%%%%%%%%%%%%%%%
\subsection{Relation between integration and trace}
Here,  we discuss the relation between the integral on $S^2$ 
and the matrix trace.

Let us consider the bundle $S\otimes L^p \otimes E$ on $S^2$ with 
the connection for $E$ vanishing.
The Dirac zero modes $\{\psi_I\}_{I=1,2,\cdots,p\times {\rm dim}E}$ for 
such bundle are written as\cite{Adachi:2020asg}
\als{
    \psi_I(z) = \psi_r^{(p)+}(z)
    \ket{+}\otimes e_i.
}
Here, $\{e_i | i=1,2, \dots, {\rm dim}E \}$ is an orthnormal basis of the 
fiber of $E$, $\ket{+}$ represents a basis vector for the positive chirality modes
of $S$, and $\psi_r^{(p)+}(z) (r=-(p-1)/2, -(p-1)/2+1, \cdots, (p-1)/2)$ 
is the function given by
\als{
    \psi_r^{(p)+}(z) = \sqrt{\frac{p}{3\pi}}\frac{1}{(1+ |z|^2)^{\frac{p-1}{2}}} 
     \qty({}_{p-1}{\rm C}_{\frac{p-1}{2}-r})^{1/2} z^{\frac{p-1}{2}-r}.
}
This is just the wave function for the Bloch coherent state \cite{Ishiki:2015saa}
and satisfies
\als{
    \sum_r \psi_r^{(p)+}(z)\psi_r^{(p)+\dagger}(z) =\frac{p}{2\pi}.
}
Then, the matrix trace for a Toeplitz operator is translated into the 
integral of functions as 
\als{
    \label{eq:trint}
    {\rm Tr}(T_p(\varphi))
    =\sum_{I}\int \omega \psi_I^\dagger \varphi \psi_I 
    = \frac{1}{4\hbar_p}\int \omega {\rm tr}\varphi.
}
Here, we have set the radius of the sphere to be 1 and hence 
$V=\frac{\pi}{2\pi}$ and $\hbar_p = \frac{V}{p} = \frac{1}{2p}$.

%%%%%%%%%%%%%%%%%%%%%%%%%%%%%%%%%%%%%%%%%%%%%%%%%%%%%%%%%%%%%
\subsection{Calculation of ${\rm Tr}\hat{X}^A\hat{X}^A$}
%%%%%%%%%%%%%%%%%%%%%%%%%%%%%%%%%%%%%%%%%%%%%%%%%%%%%%%%%%%%%
Here, we compute ${\rm Tr}\hat{X}^A\hat{X}^A$ including $1/p$ corrections.
By using the  (\ref{asym exp}) and (\ref{hatX}), we have
\als{
    {\rm Tr}\hat{X}^A\hat{X}^A
    &= \hbar_p{\rm Tr}T_p\qty(\Tilde{X}^A X^A + X^A\Tilde{X}^A)
    + \hbar^2_p{\rm Tr}T_p\qty(C_1(\Tilde{X}^A, X^A) + C_1(X^A,\Tilde{X}^A) + \Tilde{X}^A\Tilde{X}^A)\\
    &\quad\quad\quad\quad\quad\quad\
    + \hbar^3_p{\rm Tr}T_p\qty(C_2(\Tilde{X}^A ,X^A) + C_2(X^A,\Tilde{X}^A) + C_1(\Tilde{X}^A,\Tilde{X}^A)) + O(\hbar_p^4),
    \label{trxx expansion}
}
where we have ignored ${\rm Tr}\qty(T_p(X^A)T_p(X^A))$, since it 
is a constant and does not depend on $A$ or $\phi$.
By substituting (\ref{expantion of fluctuations}),
we find that the first order term in $\hbar_p$ in (\ref{trxx expansion}) is simply given 
as
\als{
    &{\rm Tr}T_p\qty(\Tilde{X}^A X^A + X^A\Tilde{X}^A)
    = 2 {\rm Tr}T_p(\phi).
}
Similarly, the second order terms in (\ref{trxx expansion}) are given as
\als{
    &{\rm Tr}T_p\qty(C_1(\Tilde{X}^A, X^A) + C_1(X^A,\Tilde{X}^A))
    = {\rm Tr}T_p(i(\nabla_-A_+ - \nabla_+ A_-) -2\phi), \\
    &{\rm Tr}T_p\qty(\Tilde{X}^A\Tilde{X}^A)
    = {\rm Tr}T_p(2A_-A_+ + \phi^2),
}
and the third order terms are given as
\als{
    &{\rm Tr}T_p\qty(C_1(\Tilde{X}^A,\Tilde{X}^A))\\
    &=- {\rm Tr}T_p\left(
    \nabla_-A_+\nabla_+A_- + \nabla_-A_-\nabla_+A_+ + A_-A_+
    + 2i(\nabla_-A_+ - \nabla_+A_-)\phi + \nabla_-\phi\nabla_+\phi + \phi^2
    \right),\\% + ({\rm Tot~der})\\
    &{\rm Tr}T_p\qty(C_2(\Tilde{X}^A ,X^A) + C_2(X^A,\Tilde{X}^A))
    = -\frac{R}{4}{\rm Tr}T_p(i(\nabla_-A_+ - \nabla_+ A_-) -2\phi) .
}
By summing up all the contributions and using \eqref{eq:trint}, we find up to total derivative terms that
\als{
    {\rm Tr}\hat{X}^A\hat{X}^A 
    &= \int \frac{\omega}{4}{\rm tr}\left[
    \qty(
    2\phi
    )
    + \hbar_p\qty(
    -2\phi + 2A_-A_+ + \phi^2
    )\right.\\
    &\left. \quad\quad\quad\quad\quad
    + \hbar^2_p\qty(
    -2\nabla_-A_+\nabla_+A_- 
    - 2i(\nabla_-A_+ - \nabla_+A_-)\phi - \nabla_-\phi\nabla_+\phi - \phi^2 +\frac{R}{2}\phi
    )
    \right] + O(\hbar_p^3)\\
    &= \int \frac{\omega}{4}{\rm tr}\left[
    \qty(2\chi) + \hbar_p(2F_{12}^0\chi -2\chi + \chi^2) + \hbar_p^2\qty(F_{12}^0\chi^2 - D_-\chi D_+\chi - \chi^2 +\frac{R}{2}\chi)
    \right]+ O(\hbar_p^3).
}
Here, 
in the second equality, we first substituted (\ref{SW expansion}) and then 
used \eqref{eq:core_chi2}, \eqref{eq:core_chi1} and \eqref{eq:core_chichi1}.
Note that the last expression is manifestly gauge invariant inheriting 
the gauge invariance of the matrix model.
Hence the result is consistent with the property of the Seiberg-Witten map.

%%%%%%%%%%%%%%%%%%%%%%%%%%%%%%%%%%%%%%%%%%%%%%%%%%%%%%%%%%%%%
\subsection{Calculation of ${\rm Tr}\epsilon_{ABC}\hat{X}^A \hat{X}^B \hat{X}^C$}
%%%%%%%%%%%%%%%%%%%%%%%%%%%%%%%%%%%%%%%%%%%%%%%%%%%%%%%%%%%%%
Here, we calculate ${\rm Tr}\epsilon_{ABC}\hat{X}^A \hat{X}^B \hat{X}^C$.
The calculation is completely parallel to the previous subsection.
By using the  (\ref{asym exp}) and (\ref{hatX}), we obtain
\als{
    {\rm Tr}(\epsilon_{ABC}\hat{X}^A &\hat{X}^B \hat{X}^C) \\
    =\epsilon_{ABC}{\rm Tr}T_p(&
     \hbar_p^2(
    X^A\Tilde{X}^B\Tilde{X}^C
    + X^A C_1(X^B, \Tilde{X}^C)
    + X^A C_1(\Tilde{X}^B, X^C)
    + \Tilde{X}^A C_1(X^B, X^C)
    )\\
    &+ \hbar_p^3(
    \Tilde{X}^A\Tilde{X}^B\Tilde{X}^C
    + X^AC_1(\Tilde{X}^B, \Tilde{X}^C)
    + \Tilde{X}^A C_1(X^B, \Tilde{X}^C)\\
    &\quad\quad
    + \Tilde{X}^A C_1(\Tilde{X}^B, X^C)
    + C_1(X^A, \Tilde{X}^B \Tilde{X}^C) 
    + X^A C_2(X^B, \Tilde{X}^C) \\
    &\quad\quad
    + X^A C_2(\Tilde{X}^B, X^C)
    + \Tilde{X}^A C_2(X^B, X^C)
    + C_1(X^A, C_1(X^B, \Tilde{X}^C))\\
    &\quad\quad
    + C_1(X^A, C_1(\Tilde{X}^B, X^C))
    + C_1(\Tilde{X}^A, C_1(X^B, X^C))
    ) \\
    &+ \hbar_p^4(
    \Tilde{X}^AC_1(\Tilde{X}^B, \Tilde{X}^C)
    + C_1(\Tilde{X}^A, \Tilde{X}^B\Tilde{X}^C)
    + C_2(X^A, \Tilde{X}^B \Tilde{X}^C) \\
    &\quad\quad
    + X^A C_2(\Tilde{X}^B, \Tilde{X}^C)
    + \Tilde{X}^A C_2(X^B, \Tilde{X}^C)
    + \Tilde{X}^AC_2(\Tilde{X}^B, X^C) \\
    &\quad\quad
    + C_1(X^A, C_1(\Tilde{X}^B, \Tilde{X}^C))
    + C_1(\Tilde{X}^A, C_1(X^B, \Tilde{X}^C))
    + C_1(\Tilde{X}^A, C_1(\Tilde{X}^B, X^C)) \\
    &\quad\quad
    + X^A C_3(X^B, \Tilde{X}^C)
    + X^A C_3(\Tilde{X}^B, X^C)
    + \Tilde{X}^A C_3(X^B, X^C) \\
    &\quad\quad
    + C_1(X^A, C_2(X^B, \Tilde{X}^C))
    + C_1(X^A, C_2(\Tilde{X}^B, X^C))
    + C_1(\Tilde{X}^A, C_2(X^B, X^C)) \\
    &\quad\quad
    + C_2(X^A, C_1(X^B, \Tilde{X}^C))
    + C_2(X^A, C_1(\Tilde{X}^B, X^C))
    + C_2(\Tilde{X}^A, C_1(X^B, X^C))
    )) + O(\hbar_p^5).
    \label{xxx expansion}
}
By using (\ref{expantion of fluctuations}), we can evaluate
the second order terms of $\hbar_p$ as
\als{
    &\epsilon_{ABC}{\rm Tr}T_p(X^A\Tilde{X}^B\Tilde{X}^C)
    = {\rm Tr}T_p(i[A_-, A_+]), \\
    &\epsilon_{ABC}{\rm Tr}T_p(X^A C_1(X^B, \Tilde{X}^C))
    = {\rm Tr}T_p(\nabla_+A_- + i\phi), \\
    &\epsilon_{ABC}{\rm Tr}T_p(X^A C_1(\Tilde{X}^B, X^C))
    = {\rm Tr}T_p(-\nabla_-A_+ + i\phi), \\
    &\epsilon_{ABC}{\rm Tr}T_p(\Tilde{X}^A C_1(X^B, X^C))
    = {\rm Tr}T_p(i\phi).
}
Taking a sum of these, we find that the second order contribution in  
    (\ref{xxx expansion}) is
\als{
    {\rm Tr}T_p(3i\phi + i[A_-, A_+] - \nabla_-A_+ + \nabla_+A_-) = 3i {\rm Tr}T_p(\phi).
}
Next, the third order terms in $\hbar_p$ in (\ref{xxx expansion}) 
are calculated as
\als{
    &\epsilon_{ABC}{\rm Tr}T_p(\Tilde{X}^A\Tilde{X}^B\Tilde{X}^C)
    = {\rm Tr}T_p(3i[A_-,A_+]\phi),\\
    &\epsilon_{ABC}{\rm Tr}T_p(C_1(X^A, \Tilde{X}^B \Tilde{X}^C))=0 ,\\
    &\epsilon_{ABC}{\rm Tr}T_p(X^AC_1(\Tilde{X}^B, \Tilde{X}^C)))
    = {\rm Tr}T_p(i\nabla_-A_+\nabla_+A_- - \nabla_-A_+\phi -i\nabla_-A_-\nabla_+A_+ + \nabla_+A_-\phi + i\phi^2),\\
    &\epsilon_{ABC}{\rm Tr}T_p(\Tilde{X}^A C_1(\Tilde{X}^B, X^C)))
    = {\rm Tr}T_p(iA_+A_- + A_+\nabla_-\phi - \phi\nabla_-A_+ + i\phi^2),\\
    &\epsilon_{ABC}{\rm Tr}T_p(\Tilde{X}^A C_1(X^B, \Tilde{X}^C)))
    = {\rm Tr}T_p(iA_-A_+ - A_-\nabla_+\phi + \phi\nabla_+A_- + i\phi^2),\\
    &\epsilon_{ABC}{\rm Tr}T_p(X^A C_2(X^B, \Tilde{X}^C))+ X^A C_2(\Tilde{X}^B, X^C))+ \Tilde{X}^A C_2(X^B, X^C))),\\
    &= {\rm Tr}T_p\qty(-\frac{R}{4}(-(\nabla_-A_+ -\nabla_+A_-) + 3i\phi)),\\
    &\epsilon_{ABC}{\rm Tr}T_p(C_1(X^A, C_1(X^B, \Tilde{X}^C)))
    ={\rm Tr}T_p(-\nabla_+A_-+ i\phi),\\
    &\epsilon_{ABC}{\rm Tr}T_p(C_1(X^A, C_1(\Tilde{X}^B, X^C)))
    ={\rm Tr}T_p(-(\nabla_-A_+ -\nabla_+A_-) - i\nabla_-\nabla_+\phi + i\phi)),\\
    &\epsilon_{ABC}{\rm Tr}T_p(C_1(\Tilde{X}^A, C_1(X^B, X^C)))
    ={\rm Tr}T_p(\nabla_-A_++ i\phi).
}
Thus, we find that the third order contribution in (\ref{xxx expansion}) is
\als{
  {\rm Tr}T_p\qty(3iF_{12}\phi + 3i\phi^2 + i\qty(2+\frac{R}{2})A_- A_+ - \qty(3+\frac{3R}{4})\phi).
}
Finally, let us consider the 4th order terms in $\hbar_p$ in (\ref{xxx expansion}).
We first consider the linear terms in $\tilde{X}$. 
Most terms are actually vanishing, since we can ignore the total derivative terms  
and $C_2(X,X)=C_3(X,X)=0$.  The only surviving contributions are 
linear terms in $\phi$, which take the following forms,
\als{
    \phi \epsilon^{ABC}X^AC_1(X^B,X^C)&=i\phi, \\
    \phi \epsilon^{ABC}C_1(X^A,C_1(X^B,X^C)
    &=\phi \epsilon^{ABC}\nabla_-X^A\nabla_+(\nabla_-X^B\nabla_+X^C)
    =-i\phi.
}
The total coefficient for this type of contributions is given as
\als{
    \qty(\frac{3R^2}{16}+\frac{6R}{4})i\phi
    =\qty(\frac{3}{4}+3)i\phi
    =\frac{15}{4}i\phi .
}
We then compute the 4th order terms in $\hbar_p$ in (\ref{xxx expansion}) 
which are quadratic or qubic in $\tilde{X}$.
By using (\ref{expantion of fluctuations}), we find that each term is given 
as follows:
\als{
    &\epsilon_{ABC}{\rm Tr}T_p(\Tilde{X}^AC_1(\Tilde{X}^B, \Tilde{X}^C))) \\
    &= -{\rm Tr}T_p(
    -iA_+\nabla_-A_-(-iA_+ + \nabla_+\phi)
    + iA_+(iA_- + \nabla_-\phi)(-\nabla_+A_- -i \phi) \\
    &\quad\quad\quad\quad
    - iA_-(-\nabla_-A_+ +i \phi)(-iA_+ + \nabla_+\phi)
    - iA_-(iA_- + \nabla_-\phi)\nabla_+A_+ \\
    &\quad\quad\quad\quad\quad\quad\quad\quad
    - i\phi(-\nabla_-A_+ +i \phi)(-\nabla_+A_- -i \phi)
    + i\phi\nabla_-A_-\nabla_+A_+
    ),
}
\als{
    &\epsilon_{ABC}{\rm Tr}T_p(C_1(\Tilde{X}^A, \Tilde{X}^B\Tilde{X}^C)) \\
    &= \epsilon_{ABC}{\rm Tr}T_p(\Tilde{X}^AC_1(\Tilde{X}^B, \Tilde{X}^C)))\\
    &\quad\quad-  {\rm Tr}T_p(
    i \nabla_-A_- A_+ (-iA_+ + \nabla_+\phi)
    + i (iA_- + \nabla_-\phi)A_+ (-\nabla_+A_- -i \phi) \\
    &\quad\quad\quad\quad\quad
    + i (-\nabla_-A_+ +i \phi) A_- (-iA_+ + \nabla_+\phi)
    +i (iA_- + \nabla_-\phi)A_- \nabla_+A_+ \\
    &\quad\quad\quad\quad\quad\quad\quad\quad\quad
    + i (-\nabla_-A_+ +i \phi)\phi (-\nabla_-A_+ +i \phi)
    - i \nabla_-A_-\phi\nabla_+A_+
    ),
}
\als{
    &\epsilon_{ABC}{\rm Tr}T_p(C_2(X^A, \Tilde{X}^B \Tilde{X}^C) + \Tilde{X}^A C_2(X^B, \Tilde{X}^C) + \Tilde{X}^AC_2(\Tilde{X}^B, X^C))\\
    &= {\rm Tr}T_p\qty(
    -\frac{iR}{2}A_-A_+ + \frac{R}{2} (\nabla_-A_+ -\nabla_+A_-)\phi - \frac{R}{2}i\phi^2
    ),
}
\als{
    &\epsilon_{ABC}{\rm Tr}T_p(X^A C_2(\Tilde{X}^B, \Tilde{X}^C)) \\
    &= {\rm Tr}T_p\left(
    -\frac{R}{4}(i\nabla_-A_+\nabla_+A_- - \nabla_-A_+\phi -i\nabla_-A_-\nabla_+A_+ + \nabla_+A_-\phi + i\phi^2)\right.\\
    &\left. \quad\quad\quad\quad\quad
    +\frac{1}{2}(-i(-\nabla_-^2A_+ -A_- + 2i\nabla_-\phi)(-\nabla_+^2A_- -A_+ - 2i\nabla_-\phi)) + i\nabla_-^2A_-\nabla_+^2A_+
    \right),
}
\als{
    &\epsilon_{ABC}{\rm Tr}T_p(C_1(\Tilde{X}^A, C_1(\Tilde{X}^B, X^C))) \\
    &= {\rm Tr}T_p(
    (-\nabla_-A_+ +i \phi)(-i\nabla_-A_+ +i\nabla_+A_- + \nabla_-\nabla_+\phi - \phi)
    - (iA_- + \nabla_-\phi)(-\nabla_+\nabla_-A_+ + i\nabla_+\phi)
    ),
}
\als{
    &\epsilon_{ABC}{\rm Tr}T_p(C_1(X^A, C_1(\Tilde{X}^B, \Tilde{X}^C))+C_1(\Tilde{X}^A, C_1(X^B, \Tilde{X}^C))) \\
    &= {\rm Tr}T_p(
    (-\nabla_-\nabla_+A_- - i\nabla_-\phi)(-iA_+ + \nabla_+\phi)
    - (-i\nabla_-A_+ +i\nabla_+A_- + \nabla_-\nabla_+\phi - \phi)(-\nabla_-A_+ +i \phi)\\
    &\quad\quad\quad\quad\quad\quad\quad\quad\quad\quad\quad\quad\quad\quad\quad\quad\quad\quad\quad\quad\quad
     - i (-\nabla_-A_+ +i \phi)(-\nabla_+A_- -i \phi) + i \nabla_-A_-\nabla_+A_+ ).
}
Summing all the above terms, we obtain the 4th order contribution in $\hbar_p$ in (\ref{xxx expansion}) as
\als{
    & {\rm Tr}T_p\left(
    \frac{15}{4}i\phi
    + i\phi^3
    - 3\nabla_-A_+A_+A_- 
    + 3A_+A_-\nabla_+A_- 
    - 3A_-\phi\nabla_+\phi 
    + A_+\nabla_-\phi\phi \right.\\
    &\quad\quad\quad\quad
    - 3i\nabla_+A_-A_+\nabla_-\phi
    - 3iA_-\nabla_-A_+\nabla_+\phi+ 3iA_+A_-\phi
    - 3iA_+A_-\nabla_+\nabla_-\phi \\
    &\quad\quad\quad\quad
    + \frac{3}{2}i(\nabla_-A_+ -\nabla_+A_-)[\nabla_+A_-,\phi]
    + \frac{3}{2}i(\nabla_-A_+ -\nabla_+A_-)[\nabla_-A_+,\phi] \\
    &\quad\quad\quad\quad
    -\frac{9}{2}i\phi^2 
    - 6i\nabla_-\phi\nabla_+\phi 
    + \frac{15}{2}i(\nabla_-A_+ -\nabla_+A_-)\phi
    + \frac{15}{2}i(\nabla_-A_+ -\nabla_+A_-)\nabla_-\nabla_+\phi \\
    &\left. \quad\quad\quad\quad
    - \frac{3}{2}iA_-A_+
    - 6i\nabla_-A_+\nabla_+A_-
    - \frac{3}{2}i\nabla_-A_+\nabla_-A_+
    - \frac{3}{2}i\nabla_+A_-\nabla_+A_-
    \right).
}
Thus, we find that 
${\rm Tr}\epsilon_{ABC}\hat{X}^A \hat{X}^B \hat{X}^C$ is given as follows.
\als{
    &{\rm Tr}\qty(\epsilon_{ABC}\hat{X}^A \hat{X}^B \hat{X}^C) \\
    =& \int \frac{\omega}{4}{\rm tr}\left(
    + 3\hbar_pi \phi
    + \hbar_p^2\qty(3iF_{12}\phi + 3i\phi^2 + i\qty(2+\frac{R}{2})A_- A_+ - \qty(3+\frac{3R}{4})\phi) \right.\\
    &\quad\quad\quad
    + \hbar_p^3\left(
    \frac{15}{4}i\phi
    + i\phi^3
    - 3\nabla_-A_+A_+A_- 
    + 3A_+A_-\nabla_+A_- 
    - 3A_-\phi\nabla_+\phi 
    + A_+\nabla_-\phi\phi \right.\\
    &\quad\quad\quad\quad\quad\quad\quad
    - 3i\nabla_+A_-A_+\nabla_-\phi
    - 3iA_-\nabla_-A_+\nabla_+\phi+ 3iA_+A_-\phi
    - 3iA_+A_-\nabla_+\nabla_-\phi \\
    &\quad\quad\quad\quad\quad\quad\quad
    + \frac{3}{2}i(\nabla_-A_+ -\nabla_+A_-)[\nabla_+A_-,\phi]
    + \frac{3}{2}i(\nabla_-A_+ -\nabla_+A_-)[\nabla_-A_+,\phi] \\
    &\quad\quad\quad\quad\quad\quad\quad
    -\frac{9}{2}i\phi^2 
    - 6i\nabla_-\phi\nabla_+\phi 
    + \frac{15}{2}i(\nabla_-A_+ -\nabla_+A_-)\phi
    + \frac{15}{2}i(\nabla_-A_+ -\nabla_+A_-)\nabla_-\nabla_+\phi \\
    &\left.\left. \quad\quad\quad\quad\quad\quad\quad
    - \frac{3}{2}iA_-A_+
    - 6i\nabla_-A_+\nabla_+A_-
    - \frac{3}{2}i\nabla_-A_+\nabla_-A_+
    - \frac{3}{2}i\nabla_+A_-\nabla_+A_-
    \right)+ O(\hbar_p^4)
    \right) \\
    =& \int \frac{\omega}{4}{\rm tr}\left(
    + 3\hbar_pi \chi
    + 3\hbar_p^2i \chi_1
    + 3\hbar_p^3i \chi_2
    + \hbar_p^2\qty(3iF^0_{12}\chi + 3i\chi^2 + i3B_- B_+ - i\frac{15}{2}\chi) \right.\\
    &+ \hbar_p^3\qty(3(iF^0_{12}\chi_1 - (\nabla_-B_{1+} -\nabla_+B_{1-} - i[B_{1-},B_+] - i[B_{1+},B_-])\chi) + 6i\chi\chi_1 + i3B_{1-} B_+ + i3B_- B_{1+} - \frac{15}{2}\chi) \\
    &+ \hbar_p^3\left(
    \frac{15}{4}i\chi
    + i\chi^3
    - 3\nabla_-B_+B_+B_- 
    + 3B_+B_-\nabla_+B_- 
    - 3B_-\chi\nabla_+\chi 
    + B_+\nabla_-\chi\chi \right.\\
    &\quad\quad\quad\quad\quad\quad\quad
    - 3i\nabla_+B_-B_+\nabla_-\chi
    - 3iB_-\nabla_-B_+\nabla_+\chi+ 3iB_+B_-\chi
    - 3iB_+B_-\nabla_+\nabla_-\chi \\
    &\quad\quad\quad\quad\quad\quad\quad
    + \frac{3}{2}i(\nabla_-B_+ -\nabla_+B_-)[\nabla_+B_-,\chi]
    + \frac{3}{2}i(\nabla_-B_+ -\nabla_+B_-)[\nabla_-B_+,\chi] \\
    &\quad\quad\quad\quad\quad\quad\quad
    -\frac{9}{2}i\chi^2 
    - 6i\nabla_-\chi\nabla_+\chi 
    + \frac{15}{2}i(\nabla_-B_+ -\nabla_+B_-)\chi
    + \frac{15}{2}i(\nabla_-B_+ -\nabla_+B_-)\nabla_-\nabla_+\chi \\
    &\left.\left. \quad\quad\quad\quad\quad\quad\quad
    - \frac{3}{2}iB_-B_+
    - 6i\nabla_-B_+\nabla_+B_-
    - \frac{3}{2}i\nabla_-B_+\nabla_-B_+
    - \frac{3}{2}i\nabla_+B_-\nabla_+B_-
    \right)+ O(\hbar_p^4)
    \right).
    \label{xxx computation}
}
The last expression is obtained by substituting  (\ref{SW expansion}).
By using \eqref{eq:core_chi2}, \eqref{eq:core_chi1}, \eqref{eq:core_chichi1},\eqref{eq:core_Fchi} and \eqref{eq:core_F_1chi}, we can further rewrite 
 (\ref{xxx computation}) as
\als{
    &\hbar_p \int \frac{\omega}{4}{\rm tr}\left[
    3i\chi
    + 3i\hbar_p\qty(\chi^2-\frac{3}{2}\chi) \right.\\
    &\left.\quad\quad\quad
    + 3i\hbar_p^2\qty(
    - \frac{3}{2}F^0_{12}\chi
    + \frac{1}{3}\chi^3
    - \frac{3}{2}\chi^2
    + \frac{5}{4}\chi
    - \frac{1}{2}(F_{12}^0)^2
    - 2D_-\chi D_+\chi
    - \frac{1}{2}F^0_{12}(D_+D_- + D_-D_+)\chi
    )
    \right]+ O(\hbar_p^4).
}
Note that the result is gauge invariant.

%%%%%%%%%%%%%%%%%%%%%%%%%%%%%%%%%%%%%%%%%%%%%%%%%%%%%%%%%%%%%
\subsection{Calculation of ${\rm Tr}[\hat{X}^A,\hat{X}^B][\hat{X}^A,\hat{X}^B]$}
%%%%%%%%%%%%%%%%%%%%%%%%%%%%%%%%%%%%%%%%%%%%%%%%%%%%%%%%%%%%%
In this subsection, we calculate ${\rm Tr}[\hat{X}^A,\hat{X}^B][\hat{X}^A,\hat{X}^B]$ 
in the leading order in $\hbar_p$.
By using the  (\ref{asym exp}) and (\ref{hatX}), we obtain
\als{
    &{\rm Tr}\qty([\hat{X}^A,\hat{X}^B][\hat{X}^A,\hat{X}^B]) \\
    =& {\rm Tr}T_p\left(
    4\hbar_p^3 D_1(\Tilde{X}^A,X^B)D_1(X^A,X^B) \right.\\
    &\quad\quad\quad
    \hbar_p^4([\Tilde{X}^A,\Tilde{X}^B][\Tilde{X}^A,\Tilde{X}^B]
    + 2 [\Tilde{X}^A,\Tilde{X}^B](D_1(X^A,\Tilde{X}^B) + D_1(\Tilde{X}^A,X^B)) \\
    &\quad\quad\quad\quad
    + 2D_1(\Tilde{X}^A,\Tilde{X}^B)D_1(X^A,X^B)
    + 2D_1(X^A,\Tilde{X}^B)D_1(X^A,\Tilde{X}^B)
    + 2D_1(X^A,\Tilde{X}^B)D_1(\Tilde{X}^A,X^B) \\
    &\left. \quad\quad\quad\quad
    + 2(D_2(\Tilde{X}^A,X^B)D_1(X^A,X^B)
    + D_1(X^A,X^B)D_1(\Tilde{X}^A,X^B))
    )+ O(\hbar_p^5)
    \right).
}
We can compute each term as follows:
\begin{align}
    &{\rm Tr}T_p\qty(D_1(\Tilde{X}^A,X^B)D_1(X^A,X^B))
    = {\rm Tr}T_p\qty(-2\phi), \nonumber\\
    &{\rm Tr}T_p\qty([\Tilde{X}^A,\Tilde{X}^B][\Tilde{X}^A,\Tilde{X}^B])
    = {\rm Tr}T_p\qty(4[\phi,A_-][\phi,A_+] -2[A_-,A_+]^2), \nonumber\\
    &{\rm Tr}T_p\qty([\Tilde{X}^A,\Tilde{X}^B](D_1(X^A,\Tilde{X}^B) + D_1(\Tilde{X}^A,X^B))) \nonumber\\
    &= {\rm Tr}T_p\qty(2i[A_-,\phi]\nabla_+\phi
    + 2i[A_+,\phi]\nabla_-\phi
    + 2i[A_-,A_+]\nabla_+A_-
    - 2i[A_-,A_+]\nabla_-A_+
    - 8[A_-,A_+]\phi), \nonumber\\
    &{\rm Tr}T_p\qty(D_1(\Tilde{X}^A,\Tilde{X}^B)D_1(X^A,X^B)) \nonumber\\
    &= {\rm Tr}T_p\qty(2(
    \nabla_-A_-\nabla_+A_+ 
    - \phi^2
    + i\nabla_+A_-\phi
    - i\nabla_-A_+\phi
    - \nabla_-A_+\nabla_+A_-
    )),\nonumber\\
    &{\rm Tr}T_p\qty(D_1(X^A,\Tilde{X}^B)D_1(X^A,\Tilde{X}^B) )\nonumber\\
    &={\rm Tr}T_p\qty( -2(
    \phi^2
    - i\nabla_+A_-\phi
    + \nabla_-\phi\nabla_+\phi
    - iA_+\nabla_-\phi
    + iA_-\nabla_+\phi
    + A_-A_+
    + \nabla_-A_-\nabla_+A_+ 
    + \nabla_-A_+\nabla_+A_- 
    )), \nonumber\\
    &{\rm Tr}T_p\qty(D_1(X^A,\Tilde{X}^B)D_1(\Tilde{X}^A,X^B))\nonumber\\
    &= {\rm Tr}T_p\qty(2\nabla_-A_-\nabla_+A_+ 
    - 2\phi^2
    + 2i\nabla_+A_-\phi
    - 2i\nabla_-A_+\phi
    + \nabla_+A_-\nabla_+A_- 
    + \nabla_-A_+\nabla_-A_+ ),\nonumber\\
    &{\rm Tr}T_p\qty(D_2(\Tilde{X}^A,X^B)D_1(X^A,X^B) + D_1(X^A,X^B)D_1(\Tilde{X}^A,X^B))
    = {\rm Tr}T_p\qty(R\phi).
\end{align}
Thus, we obtain
\begin{align}
    &{\rm Tr}\qty([\hat{X}^A,\hat{X}^B][\hat{X}^A,\hat{X}^B]) \nonumber\\
    &=\int \frac{\omega}{4}{\rm tr}\left[
    - 8\hbar_p^2\phi \right. \nonumber\\
    &\quad\quad\quad\quad
    + \hbar_p^3(
    4[\phi,A_-][\phi,A_+] -2[A_-,A_+]^2
    + 4i[A_-,\phi]\nabla_+\phi
    + 4i[A_+,\phi]\nabla_-\phi \nonumber\\
    &\quad\quad\quad\quad\quad\quad\quad
    + 4i[A_-,A_+]\nabla_+A_-
    - 4i[A_-,A_+]\nabla_-A_+
    - 16[A_-,A_+]\phi
    - 12\phi^2 \nonumber\\
    &\quad\quad\quad\quad\quad\quad\quad
    - 16i(\nabla_-A_+-\nabla_+A_-)\phi
    - 4\nabla_-A_+\nabla_+A_-
    + 2\nabla_+A_-\nabla_+A_-
    + 2\nabla_-A_+\nabla_-A_+ \nonumber\\
    &\left. \quad\quad\quad\quad\quad\quad\quad
    - 8A_-A_+
    - 4\nabla_-\phi\nabla_+\phi
    + 2R\phi
    )+ O(\hbar_p^4)
    \right]\nonumber\\
    &=\int \frac{\omega}{4}{\rm tr}\left[
    - 8\hbar_p^2\chi
    + \hbar_p^3(
    - 2(F_{12}^0)^2
    - 4D_-\chi D_+\chi
    - 12\chi^2
    + 8F_{12}^0\chi
    + 2R\chi
    ) + O(\hbar_p^4)
    \right].
\end{align}
The result is again gauge invariant.

%%%%%%%%%%%%%%%%%%%%%%%%%%%%%%%%%%%%%%%%%%%%%%%%%%%%%%%%%%%%%
%%%%%%%%%%%%%%%%%%%%%%%%%%%%%%%%%%%%%%%%%%%%%%%%%%%%%%%%%%%%%
\section{Some properties of quantized Wilson lines}
\label{Some properties of quantized Wilson lines}
%%%%%%%%%%%%%%%%%%%%%%%%%%%%%%%%%%%%%%%%%%%%%%%%%%%%%%%%%%%%%
%%%%%%%%%%%%%%%%%%%%%%%%%%%%%%%%%%%%%%%%%%%%%%%%%%%%%%%%%%%%%
In this appendix, we derive the properties \eqref{eq:pro_her} and
\eqref{eq:pro_asy} for quantized Wilson lines.

%%%%%%%%%%%%%%%%%%%%%%%%%%%%%%%%%%%%%%%%%%%%%%%%%%%%%%%%%%%%%
%%%%%%%%%%%%%%%%%%%%%%%%%%%%%%%%%%%%%%%%%%%%%%%%%%%%%%%%%%%%%
\subsection{Hermite conjugate}
\label{The hermite conjugate of the quantized Wilson line}
%%%%%%%%%%%%%%%%%%%%%%%%%%%%%%%%%%%%%%%%%%%%%%%%%%%%%%%%%%%%%
%%%%%%%%%%%%%%%%%%%%%%%%%%%%%%%%%%%%%%%%%%%%%%%%%%%%%%%%%%%%%
Here, we show \eqref{eq:pro_her}.
First, let us consider the straight path $\gamma$ from $y$ to $x$ on torus,
\als{
 \gamma(x|y):\ z(\tau) = (1-\tau)y + \tau x ,
}
where $\tau \in [0,1]$. 
We define the inverse path of $\gamma$ by $\gamma^{-1}(x|y)=\gamma(y|x)$.
Then, we can compute the Hermitian conjugate of $T_p(U_l)$ as
\als{
    (T_p(U_l)^{\dagger})_{IJ}
    &= T_p(U_l)_{JI}^{*}
    = (U_l\psi_I,\psi_J) \\
    &= \int \omega(x) {\rm tr}\qty[
    \qty(e^{ -\frac{\hbar_pl^2}{4}}\int \omega(y) {\rm P}\exp(i\int_{\gamma(x|y)} A)B_p^{(L)}(x|y)e^{\frac{l(x^1 - y^1)}{2}}\psi(y))^{\dagger}\psi_J(x)
    ] \\
    &= \int \omega(x)\omega(y) {\rm tr}\qty[
    \psi_I^{\dagger}(y)e^{ -\frac{\hbar_pl^2}{4}}{\rm P}\exp(i\int_{\gamma(y|x)} A)B_p^{(L)}(y|x)e^{\frac{-l(y^1 - x^1)}{2}}\psi_J(x)
    ] \\
    &= \int \omega(x)\omega(y) {\rm tr}\qty[\psi_I^{\dagger}(y)U_{-l}\psi_J(x)] \\
    &= T_p(U_{-l})_{IJ}.
    \label{Tuldagger = Tu-l}
}
Here, in the forth equality, we used
\als{
    B_p^{(L)\dagger}(x^1,x^2|y^1,y^2) = B_p^{(L)}(y^1,y^2|x^1,x^2) ,
}
which follows from (\ref{bergman kernel for l}).
It is easy to show that $U_l^{\dagger}=U_{-l}$. In fact, we can calculate as
\als{
    (U_l\psi',\psi) 
    &= \int \omega(x) {\rm tr}\qty[
    \qty(e^{ -\frac{\hbar_pl^2}{4}}\int \omega(y) {\rm P}\exp(i\int_{\gamma(x|y)} A)B_p^{(L)}(x|y)e^{\frac{l(x^1 - y^1)}{2}}\psi'(y))^{\dagger}\psi(x)
    ] \\
    &= \int \omega(x)\omega(y) {\rm tr}\qty[
    \psi^{'\dagger}(y)e^{ -\frac{\hbar_pl^2}{4}}{\rm P}\exp(i\int_{\gamma(y|x)} A)B_p^{(L)}(y|x)e^{\frac{-l(y^1 - x^1)}{2}}\psi(x)
    ] \\
    &= \int \omega(x)\omega(y) {\rm tr}\qty[\psi^{'\dagger}(y)U_{-l}\psi(x)] \\
    &= (\psi' , U_{-l} \psi) .
    \label{uldagger = u-l}
}
The equation \eqref{eq:pro_her} follows from
 (\ref{Tuldagger = Tu-l}) and (\ref{uldagger = u-l}).

%%%%%%%%%%%%%%%%%%%%%%%%%%%%%%%%%%%%%%%%%%%%%%%%%%%%%%%%%%%%%
%%%%%%%%%%%%%%%%%%%%%%%%%%%%%%%%%%%%%%%%%%%%%%%%%%%%%%%%%%%%%
\subsection{Asymptotic expansion}
%%%%%%%%%%%%%%%%%%%%%%%%%%%%%%%%%%%%%%%%%%%%%%%%%%%%%%%%%%%%%
%%%%%%%%%%%%%%%%%%%%%%%%%%%%%%%%%%%%%%%%%%%%%%%%%%%%%%%%%%%%%
Here, we show the equation \eqref{eq:pro_asy}.
Below, to simplify the notation, we omit the $E$-dependence (for instance
we denote $D^{(E)}$ simply by $D$).
From the definition of the Toeplitz operator (\ref{quantization map}),
we have
\als{
    T_{p}(U_{l_1})T_{p}(V_{l_2})&=\Pi U_{l_1} \Pi V_{l_2} \Pi 
    =\Pi U_{l_1} V_{l_2} \Pi - \Pi U_{l_1}(1-\Pi) V_{l_2} \Pi .
}
Let us introduce the operator 
\begin{align}
P=
\left(
\begin{array}{cc}
0 & D^- (D^+ D^- )^{-1} \\
(D^+ D^- )^{-1} D^+ & 0 \\
\end{array}
\right),
\end{align}
where $D^+$ and $D^-$ are the restrictions of the Dirac operator $D$ 
onto the positive and negative chirality modes, respectively and
the basis is chosen such that $D$ is represented as 
\begin{align}
D=
\left(
\begin{array}{cc}
0 & D^+ \\
D^- & 0 \\
\end{array}
\right).
\end{align}
Note that since ${\rm Ker}D^+ D^- =\{ 0 \}$ for sufficiently large $p$,
$(D^+ D^- )^{-1}$ is well-defined.
It is easy to see that $PD=DP$, $(DP)^2 =DP$ and
the projection onto $({\rm Ker}D)^{\perp}$
 is given as $1-\Pi= DP= D P^2 D$. 
Thus, for $\psi_I,\psi_J \in{\rm Ker} D$, we obtain
\als{
    (\psi_I,T_p(U_{l_1})T_p(V_{l_2})\psi_J)
    &=(\psi_I , T_p(U_{l_1}V_{l_2}) \psi_J)+(D(U_{-l_1}\psi_I),P^2D(V_{l_2}\psi_J)).
}
Note that, the zero mode $\psi_J$ has positive chirality for large $p$.
Since $D$ flips the chirality, $D(V_{n_1}\psi_J)$ has negative chirality.
Then, we find that
\als{
    D(V_{n_1}\psi_J) \in ({\rm Ker}D)^{\perp}.
}
Since $1-\Pi=DP$ is the identity operator on $({\rm Ker}D)^{\perp}$, 
we have $P^2=D^{-2}$ on this space. Thus, we obtain
\als{
    (\psi_I,T_p(U_{l_1})T_p(V_{l_2})\psi_J)&=(\psi_I , T_p(U_{l_1}V_{l_2}) \psi_J)+(D(U_{-l_1}\psi_I),D^{-2} D(V_{l_2}\psi_J)).
    \label{d-2}
}
Since $D$ has a large gap (namely, $D^{-2}\sim O(\hbar_p)$ on $({\rm Ker}D)^{\perp}$), 
 and both of $D(U_{-l_1}\psi_I)$ and $D(V_{n_1}\psi_J)$ are of $O(\hbar^0)$, 
the second term in (\ref{d-2}) is suppressed in the classical limit $\hbar_p \rightarrow 0$. We therefore obtain \eqref{eq:pro_asy}.

\end{appendix}

\end{document}